\documentclass[twocolumn,prx,aps,amsmath,amssymb,longbibliography]{revtex4-1}

\usepackage{graphicx}
\usepackage{dcolumn}
\usepackage{bm}

\begin{document}

\title{Majorana ensembles with fractional entropy and conductance in
  nanoscopic systems}

\author{Sergey Smirnov}
\affiliation{P. N. Lebedev Physical Institute of the Russian Academy of
  Sciences, 119991 Moscow, Russia}
\email{1) sergej.physik@gmail.com\\2)
  sergey.smirnov@physik.uni-regensburg.de\\3) ssmirnov@sci.lebedev.ru}

\date{\today}

\begin{abstract}
Quantum thermodynamics is a promising route to unambiguous detections of
Majorana bound states. Being fundamentally different from quantum transport,
this approach reveals unique Majorana thermodynamic behavior and deepens our
insight into Majorana quantum transport itself. Here we demonstrate that a
nanoscopic system with topological superconductors produces a remarkable
accumulation of Majorana thermodynamic states in wide ranges of Majorana
tunneling phases by means of increasing its temperature $T$. Revealing this
physical behavior is twofold beneficial. First, it significantly reduces the
dependence of the entropy on the tunneling phases which become almost
irrelevant in experiments. Second, the fractional Majorana entropy
$S_M^{(2)}=k_B\ln(2^\frac{3}{2})$ may be observed at high temperatures
substantially facilitating experiments. Analyzing quantum transport, we
predict that when the temperature increases, the above thermodynamic behavior
will induce an anomalous increase of the linear conductance from vanishing
values up to the unitary fractional Majorana plateau $G_M=e^2/2h$ extending to
high temperatures.
\end{abstract}


\maketitle

\section{Introduction}\label{intro}
Non-Abelian Majorana \cite{Majorana_1937} bound states (MBSs) are promising
candidates for topological fault-tolerant quantum computation
\cite{Kitaev_2001,Kitaev_2003}. While theoretically MBSs are assumed to exist
in topological superconductors (TSs)
\cite{Alicea_2012,Flensberg_2012,Sato_2016,Aguado_2017,Lutchyn_2018}, their
conclusive detection is a challenge. A possible experimental platform for
fully conclusive Majorana signatures is quantum transport and the most direct
observable is the electric conductance. Recently there have appeared critique
\cite{Yu_2021,Frolov_2021} of numerous experiments on average electric
currents to detect MBSs. Thus a crisis in measurements of the Majorana
conductance is evident. Nevertheless, the quantum transport potential remains
far from being exhausted, awaiting its soon implementation for various
detections of MBSs. For instance, shot noise
\cite{Liu_2015,Liu_2015a,Feng_2021} is a promising route because it explicitly
reveals \cite{Smirnov_2017} a fractionalization of Dirac fermions with the
integer charge $e$ into well separated MBSs with the fractional effective
charge $e/2$. Further, finite frequency noise \cite{Valentini_2016} or quantum
noise \cite{Smirnov_2019} detects MBSs via finite frequency resonances with
fractional maxima. More comprehensive experiments involve thermoelectric
currents \cite{Leijnse_2014,Ramos-Andrade_2016,Sun_2021,Wang_2021}, universal
Majorana thermoelectric noise \cite{Smirnov_2018}, Majorana thermoelectric
quantum noise at finite frequencies \cite{Smirnov_2019a}, combinations of
thermal nonequilibrium with thermodynamic properties \cite{Smirnov_2020a} and
Majorana dynamics of other observables such as magnetization
\cite{Wrzesniewski_2021}.

A fundamentally different way to unambiguously observe MBSs is quantum
thermodynamics, in particular, measurements of the entropy of a mesoscopic
system. Whereas the value of the Majorana conductance may result from other
physical phenomena totally unrelated to MBSs, measuring a fractional Majorana
entropy $S_M=k_B\ln(2^\frac{n}{2})$, where $n$ is an odd positive integer,
$n=1,3,5,\dots$, unambiguously signals \cite{Smirnov_2015} that the system
hosts well separated halves of Dirac fermions. Indeed, all topologically
trivial zero energy bound states have an integer entropy $S_n=k_B\ln(2^n)$,
where $n$ is an integer, $n=0,1,2,\dots$, and thus such states are
unambiguously filtered out as having nothing to do with MBSs. Recent
experiments and experimental proposals on the entropy in mesoscopic systems
\cite{Hartman_2018,Kleeorin_2019,Pyurbeeva_2021,Child_2021,Child_2021a}
demonstrate that Majorana entropy measurements \cite{Sela_2019} are
feasible. However, as shown in Ref. \cite{Smirnov_2021}, difficulties arise
due to the sensitivity of the entropy to tunneling phases emerging in
realistic setups with possible interference effects \cite{Gong_2021}. Although
in a particular setup one may try to prepare TSs long enough to reduce effects
of the Majorana tunneling phases, in practical applications these phases
cannot be removed and are fundamentally used for a proper control of Majorana
qubits \cite{Gau_2020,Gau_2020a}. More crucially, experimental observations of
the Majorana entropy require very low temperatures which are hard to achieve.

Here we reveal a remarkable property of a system with MBSs to accumulate
Majorana thermodynamic states (or ensemble states) in wide ranges of tunneling
phases when the system's temperature is increased. This surprisingly solves
the two difficulties above: it significantly diminishes dependence of the
entropy on Majorana tunneling phases and at the same time substantially
increases the temperatures at which the Majorana entropy may be measured in
realistic setups of practical significance. Moreover, we demonstrate that
involving more than two MBSs essentially simplifies experiments because it
fully eliminates sensitive adjustments between the tunneling phases and gate
voltage \cite{Smirnov_2021}. Crucially for quantum transport experiments, we
predict that the emergence of these Majorana thermodynamic states leads to an
anomalous behavior of the linear conductance. Specifically, instead of a
commonly expected suppression of a unitary conductance maximum to vanishing
values, increasing the system's temperature induces here a remarkable growth
of the linear conductance from vanishing values up to the unitary fractional
Majorana plateau $G_M=e^2/2h$ extending to high temperatures. Therefore,
besides the fractional Majorana entropy, such anomalous behavior of the linear
conductance is a fully conclusive signature of MBSs in quantum transport
experiments performed in the same setup. An apparent feasibility to detect
MBSs at high temperatures by means of measuring the entropy, linear
conductance or their mutual behavior in a single mesoscopic setup is very
attractive for contemporary experiments.
\begin{figure}
\includegraphics[width=8.0 cm]{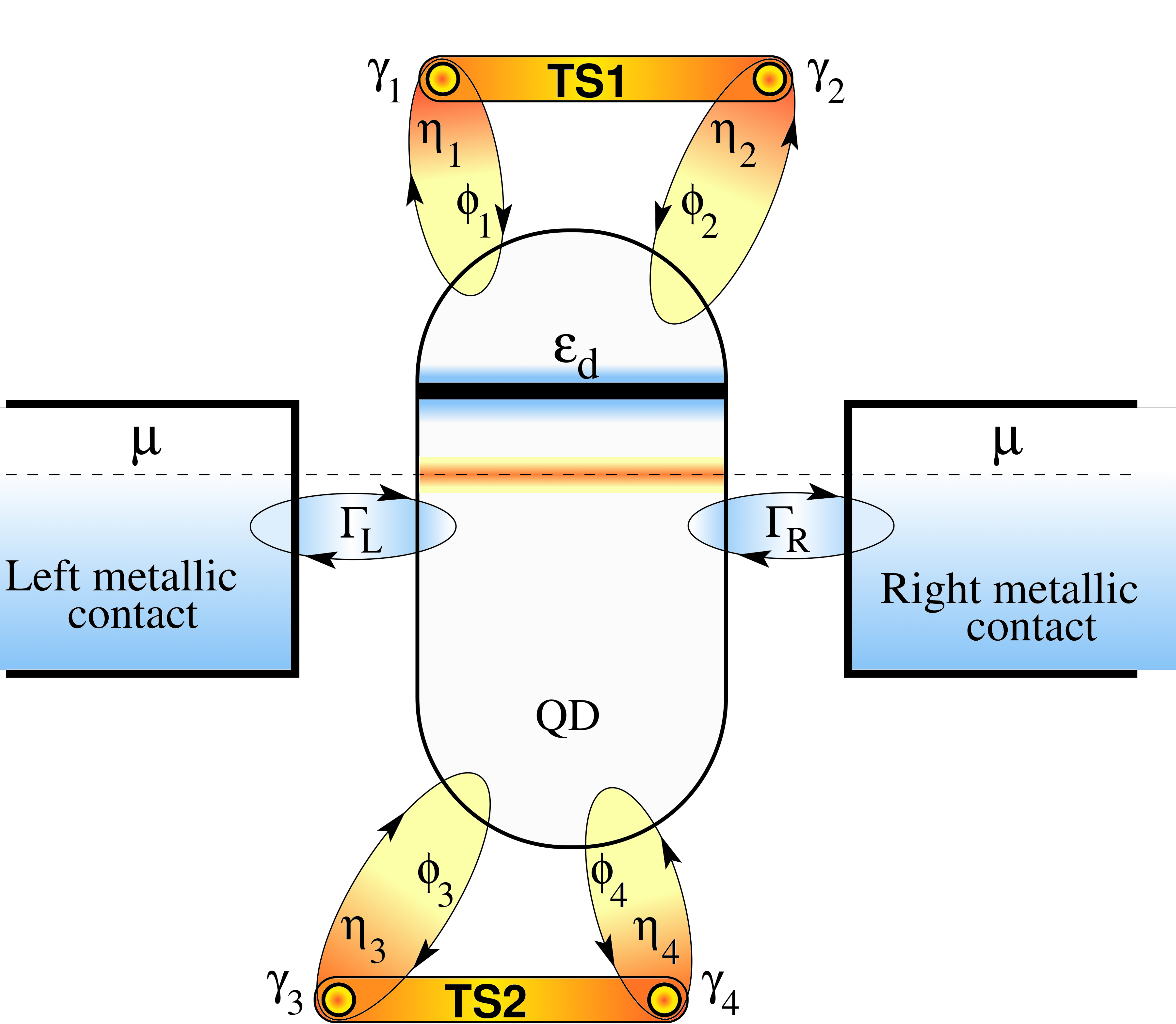}
\caption{\label{figure_1} Outline of a quantum device revealing unique
  Majorana thermodynamic and transport behavior.}
\end{figure}

The paper is organized as follows. Section \ref{mselc} describes a Majorana
setup, its theoretical model and the expressions for the entropy and linear
conductance in terms of the quantum dot Green's functions. The expressions for
the Green's functions are also provided. In Section \ref{results} we show how
the entropy depends on the tunneling phases at low and high temperatures,
demonstrate the Majorana universality of the entropy and linear conductance
and explore their temperature dependence at different values of the tunneling
phases. Finally, Section \ref{concl} concludes the paper and presents possible
outlooks.
\section{Majorana setup, entropy and linear conductance}\label{mselc}
To explore quantum thermodynamic and transport signatures of MBSs we consider
the physical setup shown in Fig. \ref{figure_1}. The device represents an
essential part of a Majorana qubit and involves a quantum dot (QD), two
metallic contacts and two grounded TSs, labeled as TS1 and TS2. Both TS1 and
TS2 support MBSs at their ends, $\gamma_{1,2}$ and $\gamma_{3,4}$,
respectively. The entanglement of the QD states with the quantum states of the
two metallic contacts and two TSs is designed and controlled via inducing
tunneling interactions indicated schematically by the corresponding ellipses
with arrows. The four MBSs $\gamma_{1,2,3,4}$ are in general all involved in
the tunneling interactions between QD and TSs with, respectively, the
tunneling amplitudes $|\eta_{1,2,3,4}|$ and phases $\phi_{1,2,3,4}$. The total
Hamiltonian of this system is
$\hat{H}=\hat{H}_D+\hat{H}_C+\hat{H}_{TS}+\hat{H}_{NT}+\hat{H}_{MT}$,
where the Hamiltonians of the QD, contacts and their tunneling coupling
describe the normal part of the system,
\begin{equation}
  \begin{split}
    &\hat{H}_D=\epsilon_dd^\dagger d,\\
    &\hat{H}_C=\sum_{l=L,R}\sum_k\epsilon_kc^\dagger_{lk}c_{lk},\\
    &\hat{H}_{NT}=\sum_{l=L,R}\mathcal{T}_l\sum_kc^\dagger_{lk}d+\text{H.c.}
  \end{split}
  \label{Ham_norm_part}
\end{equation}
  The
energy level $\epsilon_d$ may be tuned with respect to the chemical potential
by a gate voltage. In particular, using $\epsilon_d\geqslant0$ excludes
possible Kondo universality \cite{Smirnov_2011a,Smirnov_2011b,Niklas_2016} and
reveals Majorana universal behavior. The contacts' density of states $\nu_c/2$
specifies the tunneling coupling, $\Gamma\equiv\sum_{l=L,R}\Gamma_l$,
$\Gamma_l\equiv\pi\nu_c|\mathcal{T}_l|^2$. The Majorana part is
\begin{equation}
  \begin{split}
    &\hat{H}_{TS}=i(\xi_{12}\gamma_2\gamma_1+\xi_{34}\gamma_4\gamma_3)/2,\\
    &\hat{H}_{MT}=\eta_1^*d^\dagger\gamma_1+\eta_2^*d^\dagger\gamma_2+\eta_3^*d^\dagger\gamma_3+\eta_4^*d^\dagger\gamma_4+\text{H.c.},
  \end{split}
  \label{Ham_SC_part}
\end{equation}
where $\gamma^\dagger_k=\gamma_k$, $\{\gamma_k,\gamma_l\}=2\delta_{kl}$,
$\eta_k=|\eta_k|\exp(i\phi_k)$, $k,l=1,2,3,4$, and $\xi_{12}$, $\xi_{34}$ are
the energies characterizing the MBSs overlap in TS1 and TS2. Although there
are four tunneling phases, physical observables must depend only on three
tunneling phase differences, {\it e.g.}, on
\begin{equation}
  \Delta\phi_{1k}\equiv\phi_1-\phi_k,\quad k=2,3,4.
  \label{Phase_differences}
\end{equation}

It is important to note that, as mentioned in the introduction, one may
prepare TS1 and TS2 long enough so that, {\it e.g.}, $|\eta_{2,4}|$ become
sufficiently small and physical observables become independent of
$\Delta\phi_{12}$ and $\Delta\phi_{14}$. However, since the couplings
$|\eta_{1,3}|$ are finite, the phase difference $\Delta\phi_{13}$ is
fundamentally present and even necessary for quantum computing using Majorana
qubits \cite{Gau_2020,Gau_2020a} whose initialization in a proper Majorana
equilibrium state may be specified by a suitable fractional value of the
entropy tuned by $\Delta\phi_{1k}$.

The entropy $S$ is found as the derivative $S=-\partial\Omega/\partial T$ of
the system's thermodynamic potential $\Omega=-k_BT\ln Z$ over the temperature
$T$. As in Refs. \cite{Smirnov_2015,Smirnov_2021}, the thermodynamic partition
function $Z$ may be derived from a skew-symmetric imaginary time field
integral \cite{Altland_2010} over the antiperiodic Grassmann fields
corresponding to the creation and annihilation operators in $\hat{H}_D$,
$\hat{H}_C$ and $\hat{H}_{TS}$. The general expression for the entropy of a
system with TSs supporting MBSs has been derived in
Ref. \cite{Smirnov_2015}. It has the following form:
\begin{equation}
  \begin{split}
    &S=k_B\ln\biggl[\cosh\biggl(\frac{\xi_{12}}{2k_BT}\biggr)\biggr]-\frac{\xi_{12}}{2T}\tanh\biggl(\frac{\xi_{12}}{2k_BT}\biggr)\\
    &+k_B\ln\biggl[\cosh\biggl(\frac{\xi_{34}}{2k_BT}\biggr)\biggr]-\frac{\xi_{34}}{2T}\tanh\biggl(\frac{\xi_{34}}{2k_BT}\biggr)\\
    &+k_B2\ln(2)+\frac{1}{8\pi k_BT^2}\int_{-\infty}^\infty
    d\epsilon\frac{\epsilon\phi(\epsilon)}{\cosh^2\bigl(\frac{\epsilon}{2k_BT}\bigr)}.
  \end{split}
  \label{Entropy}
\end{equation}
Here $\phi(\epsilon)$ is the phase of the following complex function:
\begin{equation}
  G_{hp}^A(-\epsilon)G_{hp}^R(\epsilon)-G_{hh}^A(-\epsilon)G_{pp}^R(\epsilon)=\rho(\epsilon)\exp[i\phi(\epsilon)],
  \label{Phase_GF}
\end{equation}
where the retarded and advanced hole-particle, hole-hole, and
particle-particle Green's functions are defined as follows:
\begin{equation}
  iG_{jj'}^{R,A}(t|t')\equiv
  \pm\Theta(\pm t\mp t')\langle\{d_j(t),d_{j'}(t')\}\rangle,
  \label{GF_def}
\end{equation}
where $j,j'=\{p,h\}$ and
\begin{equation}
  d_p\equiv d^\dagger,\quad d_h\equiv d.
  \label{Def_p_h_operat}
\end{equation}

The particle-hole, particle-particle, and hole-hole retarded and advanced
Green's functions are found as in Refs. \cite{Smirnov_2015,Smirnov_2021} using
the Keldysh field integral \cite{Altland_2010}. After performing the standard
Keldysh rotation and integrating out the Grassmann fields of the normal
metallic contacts and TSs, one obtains the Green's functions as the elements
of the inverse matrix of the Keldysh effective action. The retarded Green's
functions are given by the following expressions:
\begin{equation}
  G_{ij}^R(\epsilon)=\frac{g_{ij}^R(\epsilon)}{g^R(\epsilon)},
  \label{RGF}
\end{equation}
where
\begin{widetext}
\begin{equation}
  \begin{split}
    &g_{hp}^R(\epsilon)=2\hbar\biggl\{(\epsilon^2-\xi_{12}^2)(\epsilon^2-\xi_{34}^2)\bigl[i\Gamma+2(\epsilon_d+\epsilon)\bigr]
    -4(\epsilon^2-\xi_{34}^2)\bigl[\epsilon(|\eta_1|^2+|\eta_2|^2)+2\xi_{12}|\eta_1||\eta_2|\sin(\Delta\phi_{12})\bigr]\\
    &-4(\epsilon^2-\xi_{12}^2)\bigl[\epsilon(|\eta_3|^2+|\eta_4|^2)+2\xi_{34}|\eta_3||\eta_4|\sin(\Delta\phi_{14}-\Delta\phi_{13})\bigr]\biggr\},\\
    &g_{pp}^R(\epsilon)=8\hbar\epsilon\bigl[(\eta_1^2+\eta_2^2)(\xi_{34}^2-\epsilon^2)+(\eta_3^2+\eta_4^2)(\xi_{12}^2-\epsilon^2)\bigr],\\
    &g_{hh}^R(\epsilon)=8\hbar\epsilon\bigl[(\eta_1^{*2}+\eta_2^{*2})(\xi_{34}^2-\epsilon^2)+(\eta_3^{*2}+\eta_4^{*2})(\xi_{12}^2-\epsilon^2)\bigr],\\
    &g^R(\epsilon)=-(\xi_{12}^2-\epsilon^2)(\xi_{34}^2-\epsilon^2)\bigl[4\epsilon_d^2+(\Gamma-2i\epsilon)^2\bigr]+64(\epsilon^2-\xi_{34}^2)|\eta_1|^2|\eta_2|^2\sin^2(\Delta\phi_{12})\\
    &+64(\epsilon^2-\xi_{12}^2)|\eta_3|^2|\eta_4|^2\sin^2(\Delta\phi_{14}-\Delta\phi_{13})+
    8(\xi_{34}^2-\epsilon^2)\bigl[\epsilon(i\Gamma+2\epsilon)(|\eta_1|^2+|\eta_2|^2)-4\epsilon_d\xi_{12}|\eta_1||\eta_2|\sin(\Delta\phi_{12})\bigr]\\
    &+8(\xi_{12}^2-\epsilon^2)\bigl[\epsilon(i\Gamma+2\epsilon)(|\eta_3|^2+|\eta_4|^2)-4\epsilon_d\xi_{34}|\eta_3||\eta_4|\sin(\Delta\phi_{14}-\Delta\phi_{13})\bigr]+
    32\epsilon^2(|\eta_1|^2+|\eta_2|^2)(|\eta_3|^2+|\eta_4|^2)\\
    &-128\xi_{12}\xi_{34}|\eta_1||\eta_2||\eta_3||\eta_4|\sin(\Delta\phi_{12})\sin(\Delta\phi_{14}-\Delta\phi_{13})\\
    &-32\epsilon^2\biggl\{|\eta_1|^2|\eta_3|^2\cos(2\Delta\phi_{13})+|\eta_1|^2|\eta_4|^2\cos(2\Delta\phi_{14})+
    |\eta_2|^2|\eta_3|^2\cos\bigl[2(\Delta\phi_{13}-\Delta\phi_{12})\bigr]\\
    &+|\eta_2|^2|\eta_4|^2\cos\bigl[2(\Delta\phi_{14}-\Delta\phi_{12})\bigr]\biggr\}.
  \end{split}
  \label{RGF_nm_dn}
\end{equation}
\end{widetext}

The advanced hole-particle, particle-particle, and hole-hole Green's functions
are obtained from the above retarded Green's functions using the following
relations:
\begin{equation}
  \begin{split}
    &G_{hp}^A(\epsilon)=\bigr[G_{hp}^R(\epsilon)\bigl]^*,\\
    G_{pp}^A(\epsilon)=\bigr[&G_{hh}^R(\epsilon)\bigl]^*,\quad G_{hh}^A(\epsilon)=\bigr[G_{pp}^R(\epsilon)\bigl]^*,
  \end{split}
  \label{AGF}
\end{equation}
which are derived from the definition of the Green's functions given in
Eq. (\ref{GF_def}).

To obtain the linear conductance we calculate the current $I_L$ in the left
normal metallic contact. The rigorous derivation of this current has been done
in Ref. \cite{Smirnov_2018} using the Keldysh field integral
\cite{Altland_2010} with a proper source action specified by the current
operator $\hat{I}_L$. This derivation is exact, general and does not depend on
the number of TSs coupled to a QD. It provides an exact and general expression
for the current in terms of the Green's functions whose structure (see
Eqs. (\ref{RGF})-(\ref{AGF})) already depends on the number of TSs coupled to
a QD. Since in the present setup the TSs are only coupled to the QD and are
not coupled to the normal metallic contacts, the exact derivation, fully
identical to the one in Ref. \cite{Smirnov_2018}, leads to the Meir-Wingreen
formula \cite{Meir_1992} involving only the hole-particle Green's
function. Thus, assuming $\Gamma_L=\Gamma_R$ for simplicity, the current in
the left normal metallic contact is given by the following expression:
\begin{equation}
  I_L=-\frac{e\Gamma}{4\pi\hbar^2}\int_{-\infty}^\infty
  d\epsilon[f_L(\epsilon)-f_R(\epsilon)]\text{Im}[G_{hp}^R(\epsilon)].
  \label{MW_formula}
\end{equation}
Note, had the TSs been directly coupled also to the normal metallic contacts,
the expression for the current in the left normal metallic contact would have
been different from Eq. (\ref{MW_formula}). This difference would have been
expressed in additional terms such as the direct and crossed Andreev terms as
is the case in setups \cite{Li_2020,Ren_2021,Leumer_2021} where a TS is
directly coupled to normal metallic contacts. However, when TSs do not couple
directly to normal metallic contacts, as is the case in the present setup, the
direct and crossed Andreev terms do not appear.

We emphasize once again that Eq. ({\ref{MW_formula}}) is an exact result which
does not depend on the number of TSs and how their MBSs are coupled to a
QD. The dependence on the number of TSs and on the coupling of their MBSs to a
QD is exactly taken into account in the structure of the Green's functions,
Eqs. (\ref{RGF})-(\ref{AGF}). This result is a consequence of the fact that,
as mentioned above, in the present setup the TSs are only coupled to the QD
and are not coupled to the normal metallic contacts. As a result, the current
operator $\hat{I}_L$ specifying the source action (see
Ref. \cite{Smirnov_2018} or many other papers, for example,
Ref. \cite{Liu_2015}) does not change its form when more TSs are involved to
couple via their MBSs only to the QD.

In Eq. (\ref{MW_formula}) $f_{L,R}(\epsilon)$ are the Fermi-Dirac
distributions of the left and right normal metallic contacts,
\begin{equation}
  f_{L,R}(\epsilon)=\frac{1}{\exp\bigl[\frac{\epsilon-\mu_{L,R}}{k_BT}\bigr]+1}
  \label{FD_distrb}
\end{equation}
where
\begin{equation}
  \mu_{L,R}=\mu\pm eV/2
  \label{Chem_pot}
\end{equation}
and $V$ is the bias voltage. As mentioned above,
the QD energy level $\epsilon_d$ with respect to the chemical potential $\mu$
may be tuned by a gate voltage and, as in Ref. \cite{Smirnov_2021}, values of
$\epsilon_d$ are specified with reference to $\mu$.

We also note that in Eq. (\ref{MW_formula}) one traditionally assumes that for
QD setups the density of states in the normal metallic contacts is energy
independent in the range of relevant energies \cite{Altland_2010}:
\begin{equation}
  \nu_C(\epsilon)\equiv\sum_k\delta(\epsilon-\epsilon_k)\approx\frac{1}{2}\nu_c.
  \label{Cont_dos}
\end{equation}

The linear conductance $G$ is a quantum transport observable whose behavior is
fully determined by the equilibrium state of the system. It is obtained as the
derivative of the current $I_L$ with respect to the bias voltage $V$ taken at
$V=0$,
\begin{equation}
  G\equiv\frac{\partial I_L}{\partial V}\biggl|_{V=0}.
  \label{Lin_cond}
\end{equation}

It is interesting to compare the mathematical structures of
Eqs. (\ref{Entropy}) and (\ref{MW_formula}). While the linear conductance
involves only the hole-particle and only the retarded Green's function and
only its imaginary part, the entropy involves the hole-particle, hole-hole and
particle-particle Green's functions, both retarded and advanced as well as
both their imaginary and real parts. Already this mathematical comparison
shows that the entropy scans the Majorana induced properties of the system
much more comprehensively than the linear conductance and thus Majorana
induced entropy signatures are more reliable than signatures extracted from
the linear conductance.

It is also important to note that in the present setup between the two TSs
there exists in general a Majorana supercurrent flowing through the
QD. However, it is not our goal to compute this Majorana supercurrent in the
present paper which only explores the entropy, the current in the left normal
metallic contact and how they are correlated. One may also assume that under
some conditions the Majorana supercurrent has a little impact. A derivation of
the Majorana supercurrent and analysis of these conditions is a separate
problem which will be explored in our future research (see also the outlook on
the Majorana supercurrent in Section \ref{concl}).

Using Eqs. (\ref{RGF})-(\ref{AGF}), the entropy $S$ and the linear conductance
$G$ are obtained by means of numerical calculations of the corresponding
integrals.

To obtain results we will focus below on the regime
\begin{equation}
  |\eta_1|>|\eta_2|,\quad |\eta_3|>|\eta_4|,\quad |\eta_{3,4}|>|\eta_1|
  \label{Regime}
\end{equation}
assuming that TS2 is coupled to the QD stronger than TS1 and the MBSs
$\gamma_{1,3}$ are coupled to the QD stronger than the MBSs $\gamma_{2,4}$,
respectively.
\section{Results}\label{results}
\begin{figure}
\includegraphics[width=8.0 cm]{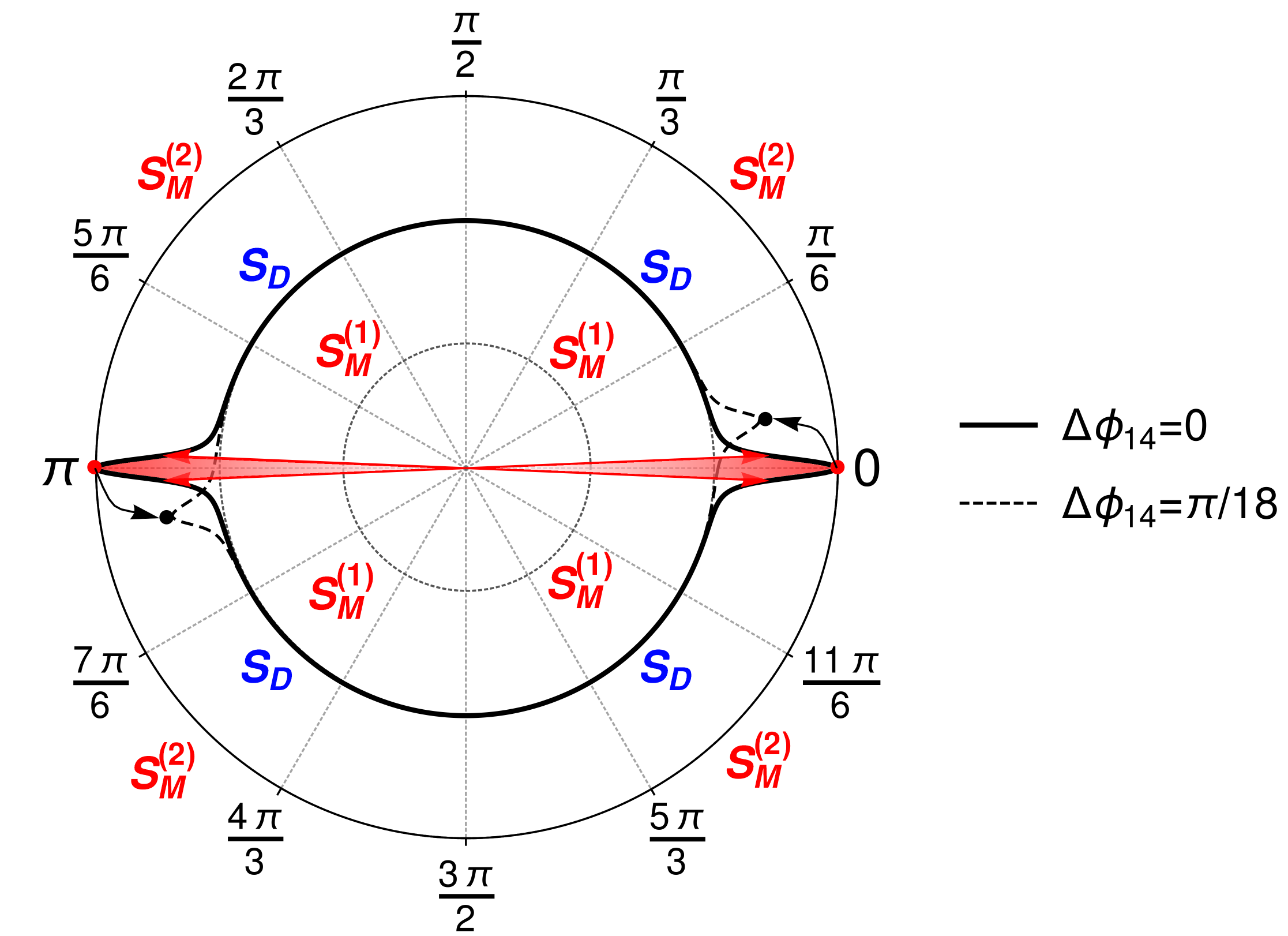}
\caption{\label{figure_2} Low temperature behavior of the entropy $S$ as a
  function of the tunneling phase difference $\Delta\phi_{13}$ in polar
  coordinates. Here $k_BT/\Gamma=10^{-5}$, $\epsilon_d/\Gamma=10^{-1}$,
  $|\eta_1|/\Gamma=10^{-2}$, $|\eta_2|/\Gamma=5\times 10^{-3}$,
  $|\eta_3|/\Gamma=1$, $|\eta_4|/\Gamma=4\times 10^{-2}$,
  $\xi_{12}/\Gamma=10^{-7}$, $\xi_{34}/\Gamma=10^{-7}$. The solid curve is for
  $\Delta\phi_{14}=0$ while the dashed curve is for
  $\Delta\phi_{14}=\pi/18$.}
\end{figure}

The dependence of the entropy on the phase difference $\Delta\phi_{13}$ in the
low temperature regime is shown in Fig. \ref{figure_2} in polar coordinates
where the distance from the center to a point on a curve is equal to the
entropy and the polar angle is equal to $\Delta\phi_{13}$. The three polar
circles correspond to the first Majorana entropy
$S_M^{(1)}=k_B\ln(2^{\frac{1}{2}})$, Dirac entropy $S_D=k_B\ln(2)$ and second
Majorana entropy $S_M^{(2)}=k_B\ln(2^{\frac{3}{2}})$. As can be seen, the
entropy has two resonances located at $\Delta\phi_{13}=\Delta\phi_{14}$ and
$\Delta\phi_{13}=\Delta\phi_{14}+\pi$. For the chosen regime,
Eq. (\ref{Regime}), the effect of $\Delta\phi_{12}$ consists in an
insignificant variation of the magnitude of the entropy resonance and thus we
put it to zero, $\Delta\phi_{12}=0$. For $\Delta\phi_{14}=0$ the maximum of
the entropy resonance is equal to $S_M^{(2)}$. We find that the width of the
entropy resonance grows when the temperature $T$ increases. In the present low
temperature regime this resonance is very narrow. Even a small deviation of
$\Delta\phi_{13}$ from $0$ or $\pi$ results in a drop of the entropy from the
Majorana value $S_M^{(2)}$ to a significantly lower one. For example, the drop
from $S_M^{(2)}$ is already about $20\%$ at the borders of the extremely
narrow angular sectors shaded in red color, where
$\Delta\phi_{13}\approx\delta,\,2\pi-\delta,\,\pi\pm\delta$ with
$\delta=0.013\pi$. Away from its resonance the entropy takes the topologically
trivial Dirac value $S_D$ which is observed almost in the whole angular range
and, as a result, it will be detected with an extremely high probability in
traditional experiments where the tunneling phases are uncontrolled. Thus we
conclude that at low temperatures topologically trivial Dirac thermodynamic
states are accumulated practically in the whole domain of the tunneling
phases. Variations of $\Delta\phi_{14}$ from zero to a finite value do not
help to induce Majorana thermodynamic states but make the situation even
worse. Indeed, the dashed curve demonstrates that $\Delta\phi_{14}$ produces
simultaneously two effects. First, it rotates the resonances from
$\Delta\phi_{13}=0$ and $\Delta\phi_{13}=\pi$ to the new locations
$\Delta\phi_{13}=\Delta\phi_{14}$ and
$\Delta\phi_{13}=\Delta\phi_{14}+\pi$. Second, it suppresses the magnitude of
the resonance from $S_M^{(2)}$ to a considerably smaller value which cannot be
increased by variations of $\Delta\phi_{12}$ as mentioned above. These two
effects are shown by the black curved arrows.

\begin{figure}
\includegraphics[width=8.0 cm]{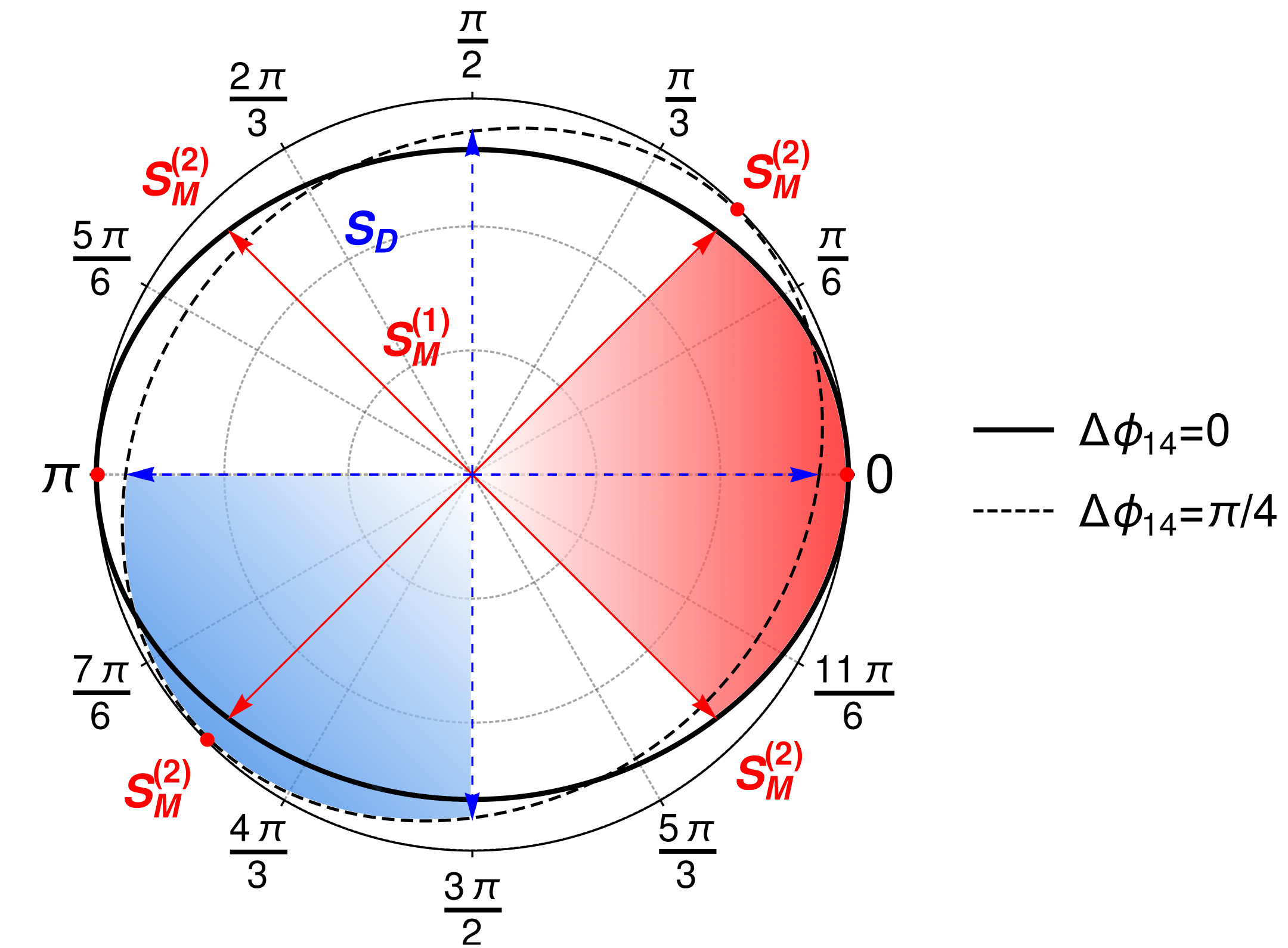}
\caption{\label{figure_3} High temperature behavior of the entropy $S$ as a
  function of the tunneling phase difference $\Delta\phi_{13}$ in polar
  coordinates. Here $k_BT/\Gamma=10^{-2}$. The other parameters have the same
  values as in Fig. \ref{figure_2}. The solid curve is for $\Delta\phi_{14}=0$
  while the dashed curve is for $\Delta\phi_{14}=\pi/4$.}
\end{figure}

The above situation is obviously unfavorable for experimental observations of
the Majorana entropy. However, it drastically changes when the temperature of
this setup increases as shown in Fig. \ref{figure_3}. As in
Fig. \ref{figure_2}, $\Delta\phi_{12}=0$. At high temperatures the entropy has
also two resonances, $\Delta\phi_{13}=\Delta\phi_{14}$,
$\Delta\phi_{13}=\Delta\phi_{14}+\pi$. However, in contrast to the low
temperature behavior, at high temperatures the entropy resonances have the
same maximum equal to $S_M^{(2)}$ for both $\Delta\phi_{14}=0$ and
$\Delta\phi_{14}=\pi/4$. Thus we conclude that at high temperatures
$\Delta\phi_{14}$ produces just a rotation of the entropy resonance. For an
experimental detection of MBSs this is favorable. Indeed, at high temperatures
$\Delta\phi_{14}$ does not ruin the Majorana entropy but just shifts it to new
values of $\Delta\phi_{13}$ without suppressing its magnitude remaining equal
to $S_M^{(2)}$. But what is much more beneficial is that due to its growth
with the temperature $T$, the width of the entropy resonance becomes very
wide. In other words, at high temperatures topologically nontrivial Majorana
thermodynamic states with the entropy $S_M^{(2)}$ are accumulated in huge
angular sectors of the tunneling phases. As a result, the Majorana entropy
$S_M^{(2)}$ may be observed in experiments with two more advantages. First, it
can be detected at very high temperatures essentially simplifying experimental
observations. Second, it can be measured in conventional experiments having no
specific control of the tunneling phases. Indeed, due to the accumulation of
topologically nontrivial Majorana thermodynamic states in vast angular
domains, the dependence of the entropy on the tunneling phases becomes almost
irrelevant. For example, the drop from the Majorana value $S_M^{(2)}$ is only
$7\%$ at the borders of the red and blue shaded angular sectors and their
$\pi$-rotated paired sectors built around the resonance centers for the solid
and dashed curves, respectively. The borders of the red shaded angular sector
and its $\pi$-rotated paired sector centered for the solid curve around,
respectively, $0$ and $\pi$ are shown by the red solid arrows,
$\Delta\phi_{13}=\pi/4,\,7\pi/4,\,3\pi/4,\,5\pi/4$. The borders of the blue
shaded angular sector and its $\pi$-rotated paired sector centered for the
dashed curve around, respectively, $5\pi/4$ and $\pi/4$ are shown by the blue
dashed arrows, $\Delta\phi_{13}=\pi,\,3\pi/2,\,0,\,\pi/2$.

\begin{figure}
\includegraphics[width=8.0 cm]{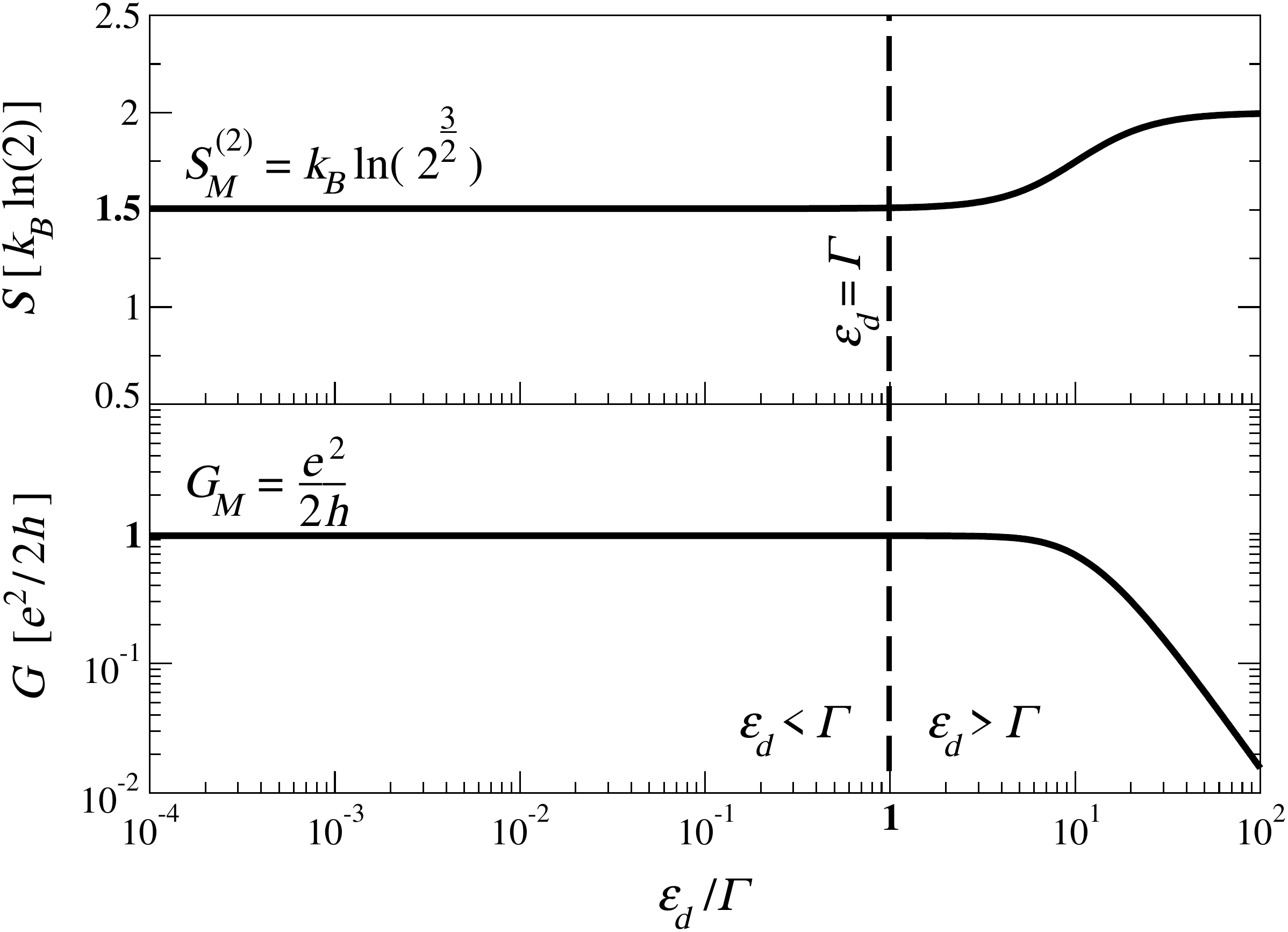}
\caption{\label{figure_4} The entropy $S$ (upper panel) and linear conductance
  $G$ (lower panel) as functions of the gate voltage controlling the location
  of the energy level $\epsilon_d$. Here $k_BT/\Gamma=10^{-2}$,
  $\Delta\phi_{13}=\pi/18$, $\Delta\phi_{14}=0$, $\Delta\phi_{12}=0$. The
  other parameters have the same values as in Fig. \ref{figure_2}.}
\end{figure}

The universality of the Majorana entropy and linear conductance is
demonstrated in Fig. \ref{figure_4}. The vertical dashed line corresponds to
$\epsilon_d=\Gamma$. On the left side, where the largest energy scale is
$\Gamma$, there emerges a universal behavior, one of MBSs' hallmarks. Indeed,
in this regime the entropy and linear conductance do not depend on the gate
voltage and remain equal to the fractional Majorana values $S_M^{(2)}$ and
$G_M$. It is important to note that the universal regime is bounded by the
contact broadening because the maximal value of $|\eta_{1,2,3,4}|$ is
$|\eta_3|=\Gamma$. If the value of $|\eta_3|$ had been much larger than
$\Gamma$, the universal regime would have been observed at gate voltages
$\epsilon_d$ much larger than $\Gamma$. The latter situation would be,
however, less realistic in experimental setups and thus we focus on the regime
where $\Gamma$ is the largest energy scale. The Majorana universality is
actually observed even up to $\epsilon_d\approx 4\Gamma$. Nevertheless,
assuming that the largest energy scale is $\Gamma$ and that
$\Gamma\approx\Delta$ ($\Delta$ is the induced superconducting gap),
increasing $\epsilon_d$ to the region $\epsilon_d>\Gamma$ might excite
quasiparticles above $\Delta$. This in turn would drive the system away from
the regime dominated by Majorana thermodynamic states and consequently screen
the effect of MBSs. However, if one only considers gate voltages with
$\epsilon_d\leqslant\Delta$, such excitations do not appear and the quantum
thermodynamic and transport properties of the system are characterized by the
universal fractional Majorana values $S_M^{(2)}$ and $G_M$. Crucially, as our
numerical calculations demonstrate, in contrast to the setup in
Ref. \cite{Smirnov_2021} with only one TS, here the second TS fully eliminates
any gate voltage dependence of the tunneling phases at which the Majorana
entropy is observed. This fact tremendously simplifies detections of the
Majorana entropy because, in contrast to the setup with only one TS considered
in Ref. \cite{Smirnov_2021}, one does not need to rely on very sensitive
adjustments between the tunneling phases and gate voltage. We have performed
many numerical calculations to cover a wide range of the tunneling amplitudes,
$10^{-3}\leqslant|\eta_{1,2,3,4}|/\Gamma\leqslant 1$, and from well separated
MBSs, $\xi_{12}/\Gamma=\xi_{34}/\Gamma=10^{-7}$ to significantly overlapping
ones, $\xi_{12}/\Gamma=\xi_{34}/\Gamma=10^{-2}$. In all the cases we have
always reproduced the universal dependence shown in Fig. \ref{figure_4}
without any gate voltage dependence of the tunneling phases at which the
Majorana entropy is observed. This demonstrates the Majorana universality also
with respect to the tunneling amplitudes $|\eta_{1,2,3,4}|$. Even when
$|\eta_1|=|\eta_2|=|\eta_3|=|\eta_4|=1$, that is outside the regime defined in
Eq. (\ref{Regime}), the Majorana entropy is observed at
$\Delta\phi_{13}=\Delta\phi_{14}=0$ which is independent of the gate
voltage. However, outside the regime defined in Eq. (\ref{Regime}) the entropy
starts to strongly depend on $\Delta\phi_{13}$. Therefore, for experimental
detections it is more favorable to couple one TS to the QD stronger than the
second one as suggested by Eq. (\ref{Regime}). This is not difficult to
achieve via the gates controlling the values of the amplitudes
$|\eta_{1,2,3,4}|$ because Eq. (\ref{Regime}) is just an inequality and not a
stringent equality and thus any fine tuning is avoided. For larger values of
$\xi_{12}/\Gamma=\xi_{34}/\Gamma>10^{-2}$ we find that the entropy is
significantly reduced and goes to very small values when
$\xi_{12}=\xi_{34}=\Gamma$.

\begin{figure}
\includegraphics[width=8.0 cm]{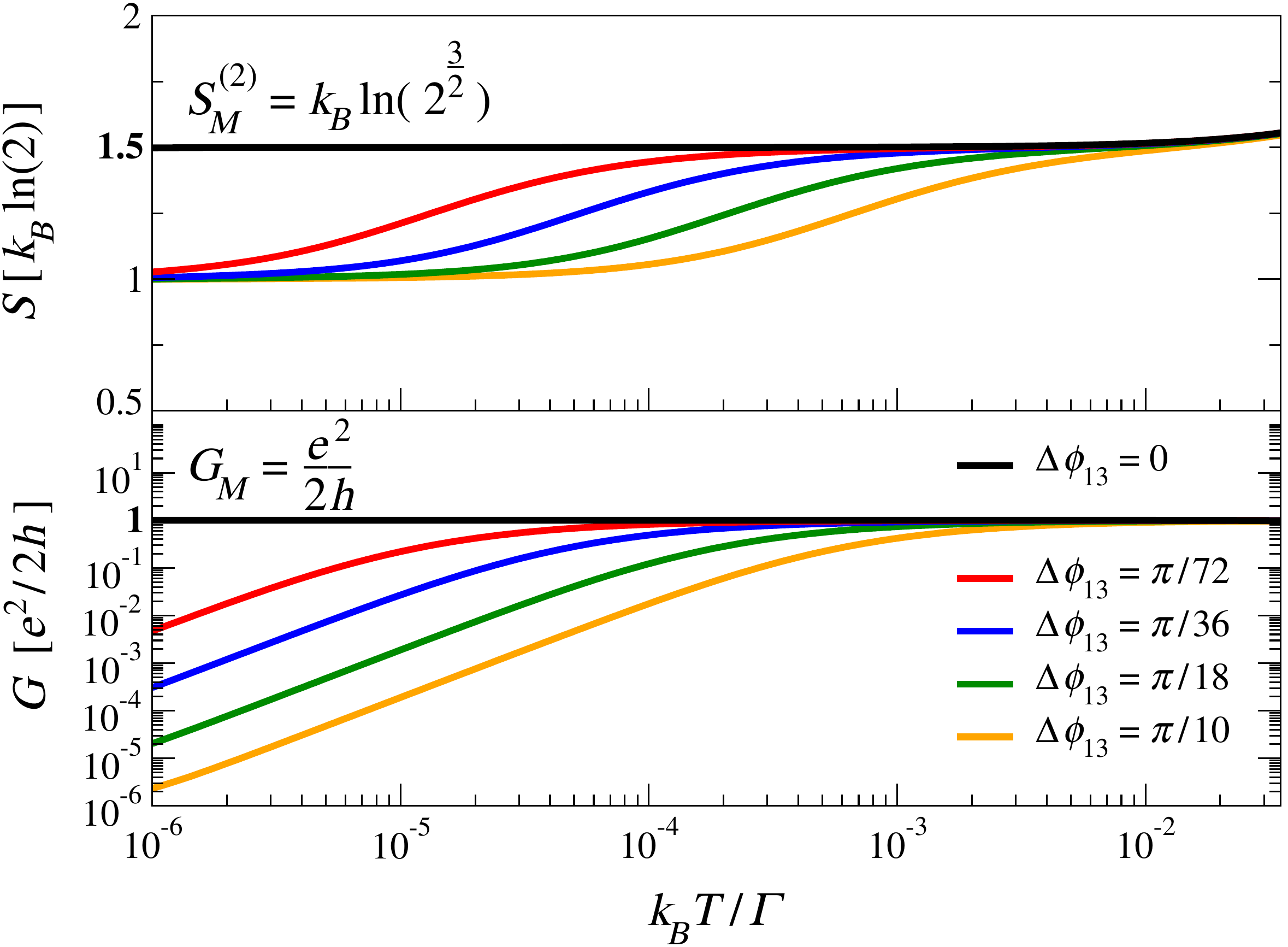}
\caption{\label{figure_5} The entropy $S$ (upper panel) and linear conductance
  $G$ (lower panel) as functions of the temperature $T$ for various values of
  the tunneling phase difference $\Delta\phi_{13}$. Here $\Delta\phi_{14}=0$,
  $\Delta\phi_{12}=0$ and the values of the other parameters are the same as
  in Fig. \ref{figure_2}.}
\end{figure}

Fig. \ref{figure_5} shows the temperature dependence of the entropy and linear
conductance. In agreement with Fig. \ref{figure_2}, at low temperatures the
entropy (upper panel) is equal to the topologically trivial Dirac value
$S_D$ for the four finite values of $\Delta\phi_{13}$ while for
$\Delta\phi_{13}=0$ it takes the topologically nontrivial Majorana value
$S_M^{(2)}$. When $T$ increases, the entropy for the finite values of
$\Delta\phi_{13}$ quickly grows from $S_D$ to the topologically nontrivial
Majorana plateau $S_M^{(2)}$ extending to very high temperatures. In
particular, at $k_BT/\Gamma=3\times 10^{-2}$ the deviation of the entropy from
the fractional plateau $S_M^{(2)}$ is $2.5\%$ for $\Delta\phi_{13}=\pi/10$. In
Ref. \cite{Mourik_2012} one finds $\Delta\approx 250\,\mu eV$ and thus the
temperature $T\approx 0.1$ K ($\Gamma\approx\Delta$, see above) which is very
high and easily achievable in modern labs. For comparison, in the setup of
Ref. \cite{Smirnov_2021} with only one TS the Majorana entropy
$S_M^{(1)}$ is observed with similar accuracy at $T\approx 30$ mK as the upper
limit meaning that in an experiment one could go to even lower temperatures
which would be hard to achieve. The above quantum thermodynamic behavior
induces striking quantum transport properties. Indeed, in accordance with
common expectations, increasing $T$ should ruin a unitary conductance maximum
down to vanishing values. Here the situation is just the
opposite. When $T$ increases, there develops an accumulation of topologically
nontrivial quantum thermodynamic states with the fractional Majorana entropy
$S_M^{(2)}$ at finite values of $\Delta\phi_{13}$. As a result, since the
linear conductance is determined by these Majorana equilibrium states, via
increasing $T$ one enhances the linear conductance (lower panel) from
vanishing values at low temperatures up to the unitary fractional Majorana
plateau $G_M$ extending to very high temperatures. For example, at
$k_BT/\Gamma=3\times 10^{-2}$ ($T\approx 0.1$ K) the deviation of the linear
conductance from the fractional Majorana plateau $G_M$ is $4\%$ for
$\Delta\phi_{13}=\pi/10$. At higher temperatures the deviations from the
Majorana plateaus will grow as can be already seen in the upper panel.
\section{Conclusion}\label{concl}
We have proposed a quantum thermodynamic approach to unambiguously detect MBSs
in nanoscale systems with TSs by means of the fractional Majorana
entropy. Specifically, we have revealed a remarkable phenomenon of an
accumulation of Majorana thermodynamic states in wide ranges of tunneling
phases via increasing the temperature $T$. This efficiently overcomes the
problem of the low temperatures at which the Majorana entropy is detected and
the problem of the sensitivity of the entropy to the tunneling
phases. Moreover, involving more than two MBSs we have fully eliminated any
gate voltage dependence of the Majorana entropy. As a result, the entropy
becomes almost independent of the tunneling phases, does not require any fine
tuning between the tunneling phases and gate voltage and its fractional
Majorana value $S_M^{(2)}=k_B\ln(2^\frac{3}{2})$ may be detected at very high
temperatures, $T\approx 0.1$ K. Since all topologically trivial zero energy
bound states have integer values of the entropy, $S_n=k_B\ln(2^n)$,
$n=0,1,2,\dots$, a detection of the fractional entropy $S_M^{(2)}$ is a fully
conclusive signature of MBSs. It has also been demonstrated how this
outstanding Majorana thermodynamic behavior induces exceptional quantum
transport properties. In particular, when $T$ increases, we have predicted an
anomalous growth of the linear conductance up to the unitary fractional
Majorana plateau $G_M=e^2/2h$. This shows how one can explore quantum
transport using a quantum thermodynamic analysis of MBSs. Similarly, one may
perform an entropy based analysis of many other equilibrium properties
characterizing the system presented in this paper. For example, one may
consider the Majorana supercurrent which is in general induced between the two
TSs and flows through the QD. Since both the entropy and supercurrent are well
defined in equilibrium, an entropic analysis of the supercurrent is a well
defined problem. Additionally, how this Majorana supercurrent is correlated
with the current in the left normal metallic contact, that is with the current
$I_L$ explored in this paper, is another important and interesting topic for
future research on other equilibrium properties of the present setup and its
generalizations to devices involving spinful quantum dots or more than two TSs
modeled via more realistic Rashba nanowires.
\section*{Acknowledgments}
The author thanks Reinhold Egger and Sergey Frolov for valuable discussions.


\begin{thebibliography}{44}%
\makeatletter
\providecommand \@ifxundefined [1]{%
 \@ifx{#1\undefined}
}%
\providecommand \@ifnum [1]{%
 \ifnum #1\expandafter \@firstoftwo
 \else \expandafter \@secondoftwo
 \fi
}%
\providecommand \@ifx [1]{%
 \ifx #1\expandafter \@firstoftwo
 \else \expandafter \@secondoftwo
 \fi
}%
\providecommand \natexlab [1]{#1}%
\providecommand \enquote  [1]{``#1''}%
\providecommand \bibnamefont  [1]{#1}%
\providecommand \bibfnamefont [1]{#1}%
\providecommand \citenamefont [1]{#1}%
\providecommand \href@noop [0]{\@secondoftwo}%
\providecommand \href [0]{\begingroup \@sanitize@url \@href}%
\providecommand \@href[1]{\@@startlink{#1}\@@href}%
\providecommand \@@href[1]{\endgroup#1\@@endlink}%
\providecommand \@sanitize@url [0]{\catcode `\\12\catcode `\$12\catcode
  `\&12\catcode `\#12\catcode `\^12\catcode `\_12\catcode `\%12\relax}%
\providecommand \@@startlink[1]{}%
\providecommand \@@endlink[0]{}%
\providecommand \url  [0]{\begingroup\@sanitize@url \@url }%
\providecommand \@url [1]{\endgroup\@href {#1}{\urlprefix }}%
\providecommand \urlprefix  [0]{URL }%
\providecommand \Eprint [0]{\href }%
\providecommand \doibase [0]{http://dx.doi.org/}%
\providecommand \selectlanguage [0]{\@gobble}%
\providecommand \bibinfo  [0]{\@secondoftwo}%
\providecommand \bibfield  [0]{\@secondoftwo}%
\providecommand \translation [1]{[#1]}%
\providecommand \BibitemOpen [0]{}%
\providecommand \bibitemStop [0]{}%
\providecommand \bibitemNoStop [0]{.\EOS\space}%
\providecommand \EOS [0]{\spacefactor3000\relax}%
\providecommand \BibitemShut  [1]{\csname bibitem#1\endcsname}%
\let\auto@bib@innerbib\@empty
\bibitem [{\citenamefont {Majorana}(1937)}]{Majorana_1937}%
  \BibitemOpen
  \bibfield  {author} {\bibinfo {author} {\bibfnamefont {E.}~\bibnamefont
  {Majorana}},\ }\bibfield  {title} {\enquote {\bibinfo {title} {Teoria
  simmetrica dell'elettrone e del positrone},}\ }\href@noop {} {\bibfield
  {journal} {\bibinfo  {journal} {Nuovo Cimento}\ }\textbf {\bibinfo {volume}
  {14}},\ \bibinfo {pages} {171} (\bibinfo {year} {1937})}\BibitemShut
  {NoStop}%
\bibitem [{\citenamefont {\text{Yu.} Kitaev}(2001)}]{Kitaev_2001}%
  \BibitemOpen
  \bibfield  {author} {\bibinfo {author} {\bibfnamefont {A.}~\bibnamefont
  {\text{Yu.} Kitaev}},\ }\bibfield  {title} {\enquote {\bibinfo {title}
  {Unpaired {M}ajorana fermions in quantum wires},}\ }\href@noop {} {\bibfield
  {journal} {\bibinfo  {journal} {Phys.-Usp.}\ }\textbf {\bibinfo {volume}
  {44}},\ \bibinfo {pages} {131} (\bibinfo {year} {2001})}\BibitemShut
  {NoStop}%
\bibitem [{\citenamefont {\text{Yu.} Kitaev}(2003)}]{Kitaev_2003}%
  \BibitemOpen
  \bibfield  {author} {\bibinfo {author} {\bibfnamefont {A.}~\bibnamefont
  {\text{Yu.} Kitaev}},\ }\bibfield  {title} {\enquote {\bibinfo {title}
  {Fault-tolerant quantum computation by anyons},}\ }\href@noop {} {\bibfield
  {journal} {\bibinfo  {journal} {Ann. Phys.}\ }\textbf {\bibinfo {volume}
  {303}},\ \bibinfo {pages} {2} (\bibinfo {year} {2003})}\BibitemShut {NoStop}%
\bibitem [{\citenamefont {Alicea}(2012)}]{Alicea_2012}%
  \BibitemOpen
  \bibfield  {author} {\bibinfo {author} {\bibfnamefont {J.}~\bibnamefont
  {Alicea}},\ }\bibfield  {title} {\enquote {\bibinfo {title} {New directions
  in the pursuit of {M}ajorana fermions in solid state systems},}\ }\href@noop
  {} {\bibfield  {journal} {\bibinfo  {journal} {Rep. Prog. Phys.}\ }\textbf
  {\bibinfo {volume} {75}},\ \bibinfo {pages} {076501} (\bibinfo {year}
  {2012})}\BibitemShut {NoStop}%
\bibitem [{\citenamefont {Leijnse}\ and\ \citenamefont
  {Flensberg}(2012)}]{Flensberg_2012}%
  \BibitemOpen
  \bibfield  {author} {\bibinfo {author} {\bibfnamefont {M.}~\bibnamefont
  {Leijnse}}\ and\ \bibinfo {author} {\bibfnamefont {K.}~\bibnamefont
  {Flensberg}},\ }\bibfield  {title} {\enquote {\bibinfo {title} {Introduction
  to topological superconductivity and {M}ajorana fermions},}\ }\href@noop {}
  {\bibfield  {journal} {\bibinfo  {journal} {Semicond. Sci. Technol.}\
  }\textbf {\bibinfo {volume} {27}},\ \bibinfo {pages} {124003} (\bibinfo
  {year} {2012})}\BibitemShut {NoStop}%
\bibitem [{\citenamefont {Sato}\ and\ \citenamefont
  {Fujimoto}(2016)}]{Sato_2016}%
  \BibitemOpen
  \bibfield  {author} {\bibinfo {author} {\bibfnamefont {M.}~\bibnamefont
  {Sato}}\ and\ \bibinfo {author} {\bibfnamefont {S.}~\bibnamefont
  {Fujimoto}},\ }\bibfield  {title} {\enquote {\bibinfo {title} {Majorana
  fermions and topology in superconductors},}\ }\href@noop {} {\bibfield
  {journal} {\bibinfo  {journal} {J. Phys. Soc. Japan}\ }\textbf {\bibinfo
  {volume} {85}},\ \bibinfo {pages} {072001} (\bibinfo {year}
  {2016})}\BibitemShut {NoStop}%
\bibitem [{\citenamefont {Aguado}(2017)}]{Aguado_2017}%
  \BibitemOpen
  \bibfield  {author} {\bibinfo {author} {\bibfnamefont {R.}~\bibnamefont
  {Aguado}},\ }\bibfield  {title} {\enquote {\bibinfo {title} {Majorana
  quasiparticles in condensed matter},}\ }\href@noop {} {\bibfield  {journal}
  {\bibinfo  {journal} {La Rivista del Nuovo Cimento}\ }\textbf {\bibinfo
  {volume} {40}},\ \bibinfo {pages} {523} (\bibinfo {year} {2017})}\BibitemShut
  {NoStop}%
\bibitem [{\citenamefont {Lutchyn}\ \emph {et~al.}(2018)\citenamefont
  {Lutchyn}, \citenamefont {Bakkers}, \citenamefont {Kouwenhoven},
  \citenamefont {Krogstrup}, \citenamefont {Marcus},\ and\ \citenamefont
  {Oreg}}]{Lutchyn_2018}%
  \BibitemOpen
  \bibfield  {author} {\bibinfo {author} {\bibfnamefont {R.~M.}\ \bibnamefont
  {Lutchyn}}, \bibinfo {author} {\bibfnamefont {E.~P. A.~M.}\ \bibnamefont
  {Bakkers}}, \bibinfo {author} {\bibfnamefont {L.~P.}\ \bibnamefont
  {Kouwenhoven}}, \bibinfo {author} {\bibfnamefont {P.}~\bibnamefont
  {Krogstrup}}, \bibinfo {author} {\bibfnamefont {C.~M.}\ \bibnamefont
  {Marcus}}, \ and\ \bibinfo {author} {\bibfnamefont {Y.}~\bibnamefont
  {Oreg}},\ }\bibfield  {title} {\enquote {\bibinfo {title} {Majorana zero
  modes in superconductor-semiconductor heterostructures},}\ }\href@noop {}
  {\bibfield  {journal} {\bibinfo  {journal} {Nat. Rev. Mater.}\ }\textbf
  {\bibinfo {volume} {3}},\ \bibinfo {pages} {52} (\bibinfo {year}
  {2018})}\BibitemShut {NoStop}%
\bibitem [{\citenamefont {Yu}\ \emph {et~al.}(2021)\citenamefont {Yu},
  \citenamefont {Chen}, \citenamefont {Gomanko}, \citenamefont {Badawy},
  \citenamefont {Bakkers}, \citenamefont {Zuo}, \citenamefont {Mourik},\ and\
  \citenamefont {Frolov}}]{Yu_2021}%
  \BibitemOpen
  \bibfield  {author} {\bibinfo {author} {\bibfnamefont {P.}~\bibnamefont
  {Yu}}, \bibinfo {author} {\bibfnamefont {J.}~\bibnamefont {Chen}}, \bibinfo
  {author} {\bibfnamefont {M.}~\bibnamefont {Gomanko}}, \bibinfo {author}
  {\bibfnamefont {G.}~\bibnamefont {Badawy}}, \bibinfo {author} {\bibfnamefont
  {E.~P. A.~M.}\ \bibnamefont {Bakkers}}, \bibinfo {author} {\bibfnamefont
  {K.}~\bibnamefont {Zuo}}, \bibinfo {author} {\bibfnamefont {V.}~\bibnamefont
  {Mourik}}, \ and\ \bibinfo {author} {\bibfnamefont {S.~M.}\ \bibnamefont
  {Frolov}},\ }\bibfield  {title} {\enquote {\bibinfo {title} {Non-{M}ajorana
  states yield nearly quantized conductance in proximatized nanowires},}\
  }\href@noop {} {\bibfield  {journal} {\bibinfo  {journal} {Nat. Phys.}\
  }\textbf {\bibinfo {volume} {17}},\ \bibinfo {pages} {482} (\bibinfo {year}
  {2021})}\BibitemShut {NoStop}%
\bibitem [{\citenamefont {Frolov}(2021)}]{Frolov_2021}%
  \BibitemOpen
  \bibfield  {author} {\bibinfo {author} {\bibfnamefont {S.}~\bibnamefont
  {Frolov}},\ }\bibfield  {title} {\enquote {\bibinfo {title} {Quantum
  computing's reproducibility crisis: {M}ajorana fermions},}\ }\href@noop {}
  {\bibfield  {journal} {\bibinfo  {journal} {Nature}\ }\textbf {\bibinfo
  {volume} {592}},\ \bibinfo {pages} {350} (\bibinfo {year}
  {2021})}\BibitemShut {NoStop}%
\bibitem [{\citenamefont {Liu}\ \emph {et~al.}(2015{\natexlab{a}})\citenamefont
  {Liu}, \citenamefont {Cheng},\ and\ \citenamefont {Lutchyn}}]{Liu_2015}%
  \BibitemOpen
  \bibfield  {author} {\bibinfo {author} {\bibfnamefont {D.~E.}\ \bibnamefont
  {Liu}}, \bibinfo {author} {\bibfnamefont {M.}~\bibnamefont {Cheng}}, \ and\
  \bibinfo {author} {\bibfnamefont {R.~M.}\ \bibnamefont {Lutchyn}},\
  }\bibfield  {title} {\enquote {\bibinfo {title} {Probing {M}ajorana physics
  in quantum-dot shot-noise experiments},}\ }\href@noop {} {\bibfield
  {journal} {\bibinfo  {journal} {Phys.\ Rev.\ B}\ }\textbf {\bibinfo {volume}
  {91}},\ \bibinfo {pages} {081405(R)} (\bibinfo {year}
  {2015}{\natexlab{a}})}\BibitemShut {NoStop}%
\bibitem [{\citenamefont {Liu}\ \emph {et~al.}(2015{\natexlab{b}})\citenamefont
  {Liu}, \citenamefont {Levchenko},\ and\ \citenamefont {Lutchyn}}]{Liu_2015a}%
  \BibitemOpen
  \bibfield  {author} {\bibinfo {author} {\bibfnamefont {D.~E.}\ \bibnamefont
  {Liu}}, \bibinfo {author} {\bibfnamefont {A.}~\bibnamefont {Levchenko}}, \
  and\ \bibinfo {author} {\bibfnamefont {R.~M.}\ \bibnamefont {Lutchyn}},\
  }\bibfield  {title} {\enquote {\bibinfo {title} {Majorana zero modes choose
  {E}uler numbers as revealed by full counting statistics},}\ }\href@noop {}
  {\bibfield  {journal} {\bibinfo  {journal} {Phys.\ Rev.\ B}\ }\textbf
  {\bibinfo {volume} {92}},\ \bibinfo {pages} {205422} (\bibinfo {year}
  {2015}{\natexlab{b}})}\BibitemShut {NoStop}%
\bibitem [{\citenamefont {Feng}\ and\ \citenamefont {Zhang}(2021)}]{Feng_2021}%
  \BibitemOpen
  \bibfield  {author} {\bibinfo {author} {\bibfnamefont {G.-H.}\ \bibnamefont
  {Feng}}\ and\ \bibinfo {author} {\bibfnamefont {H.-H.}\ \bibnamefont
  {Zhang}},\ }\bibfield  {title} {\enquote {\bibinfo {title} {Probing robust
  {M}ajorana signatures by crossed {A}ndreev reflection with a quantum dot},}\
  }\href@noop {} {\bibfield  {journal} {\bibinfo  {journal}
  {arXiv:2105.02830v2}\ } (\bibinfo {year} {2021})}\BibitemShut {NoStop}%
\bibitem [{\citenamefont {Smirnov}(2017)}]{Smirnov_2017}%
  \BibitemOpen
  \bibfield  {author} {\bibinfo {author} {\bibfnamefont {S.}~\bibnamefont
  {Smirnov}},\ }\bibfield  {title} {\enquote {\bibinfo {title} {Non-equilibrium
  {M}ajorana fluctuations},}\ }\href@noop {} {\bibfield  {journal} {\bibinfo
  {journal} {New J. Phys.}\ }\textbf {\bibinfo {volume} {19}},\ \bibinfo
  {pages} {063020} (\bibinfo {year} {2017})}\BibitemShut {NoStop}%
\bibitem [{\citenamefont {Valentini}\ \emph {et~al.}(2016)\citenamefont
  {Valentini}, \citenamefont {Governale}, \citenamefont {Fazio},\ and\
  \citenamefont {Taddei}}]{Valentini_2016}%
  \BibitemOpen
  \bibfield  {author} {\bibinfo {author} {\bibfnamefont {S.}~\bibnamefont
  {Valentini}}, \bibinfo {author} {\bibfnamefont {M.}~\bibnamefont
  {Governale}}, \bibinfo {author} {\bibfnamefont {R.}~\bibnamefont {Fazio}}, \
  and\ \bibinfo {author} {\bibfnamefont {F.}~\bibnamefont {Taddei}},\
  }\bibfield  {title} {\enquote {\bibinfo {title} {Finite-frequency noise in a
  topological superconducting wire},}\ }\href@noop {} {\bibfield  {journal}
  {\bibinfo  {journal} {Physica E}\ }\textbf {\bibinfo {volume} {75}},\
  \bibinfo {pages} {15} (\bibinfo {year} {2016})}\BibitemShut {NoStop}%
\bibitem [{\citenamefont {Smirnov}(2019{\natexlab{a}})}]{Smirnov_2019}%
  \BibitemOpen
  \bibfield  {author} {\bibinfo {author} {\bibfnamefont {S.}~\bibnamefont
  {Smirnov}},\ }\bibfield  {title} {\enquote {\bibinfo {title} {Majorana
  finite-frequency nonequilibrium quantum noise},}\ }\href@noop {} {\bibfield
  {journal} {\bibinfo  {journal} {Phys.\ Rev.\ B}\ }\textbf {\bibinfo {volume}
  {99}},\ \bibinfo {pages} {165427} (\bibinfo {year}
  {2019}{\natexlab{a}})}\BibitemShut {NoStop}%
\bibitem [{\citenamefont {Leijnse}(2014)}]{Leijnse_2014}%
  \BibitemOpen
  \bibfield  {author} {\bibinfo {author} {\bibfnamefont {M.}~\bibnamefont
  {Leijnse}},\ }\bibfield  {title} {\enquote {\bibinfo {title} {Thermoelectric
  signatures of a {M}ajorana bound state coupled to a quantum dot},}\
  }\href@noop {} {\bibfield  {journal} {\bibinfo  {journal} {New J. Phys.}\
  }\textbf {\bibinfo {volume} {16}},\ \bibinfo {pages} {015029} (\bibinfo
  {year} {2014})}\BibitemShut {NoStop}%
\bibitem [{\citenamefont {Ramos-Andrade}\ \emph {et~al.}(2016)\citenamefont
  {Ramos-Andrade}, \citenamefont {{\'Avalos-Ovando}}, \citenamefont
  {Orellana},\ and\ \citenamefont {Ulloa}}]{Ramos-Andrade_2016}%
  \BibitemOpen
  \bibfield  {author} {\bibinfo {author} {\bibfnamefont {J.~P.}\ \bibnamefont
  {Ramos-Andrade}}, \bibinfo {author} {\bibfnamefont {O.}~\bibnamefont
  {{\'Avalos-Ovando}}}, \bibinfo {author} {\bibfnamefont {P.~A.}\ \bibnamefont
  {Orellana}}, \ and\ \bibinfo {author} {\bibfnamefont {S.~E.}\ \bibnamefont
  {Ulloa}},\ }\bibfield  {title} {\enquote {\bibinfo {title} {Thermoelectric
  transport through {M}ajorana bound states and violation of
  {W}iedemann-{F}ranz law},}\ }\href@noop {} {\bibfield  {journal} {\bibinfo
  {journal} {Phys.\ Rev.\ B}\ }\textbf {\bibinfo {volume} {94}},\ \bibinfo
  {pages} {155436} (\bibinfo {year} {2016})}\BibitemShut {NoStop}%
\bibitem [{\citenamefont {Sun}\ and\ \citenamefont {Chi}(2021)}]{Sun_2021}%
  \BibitemOpen
  \bibfield  {author} {\bibinfo {author} {\bibfnamefont {L.-L.}\ \bibnamefont
  {Sun}}\ and\ \bibinfo {author} {\bibfnamefont {F.}~\bibnamefont {Chi}},\
  }\bibfield  {title} {\enquote {\bibinfo {title} {Detecting spin heat
  accumulation by sign reversion of thermopower in a quantum dot side‐coupled
  to {M}ajorana bound states},}\ }\href@noop {} {\bibfield  {journal} {\bibinfo
   {journal} {J. Low Temp. Phys.}\ } (\bibinfo {year} {2021})}\BibitemShut
  {NoStop}%
\bibitem [{\citenamefont {Wang}\ and\ \citenamefont {Huang}(2021)}]{Wang_2021}%
  \BibitemOpen
  \bibfield  {author} {\bibinfo {author} {\bibfnamefont {Z.-H.}\ \bibnamefont
  {Wang}}\ and\ \bibinfo {author} {\bibfnamefont {W.-C.}\ \bibnamefont
  {Huang}},\ }\bibfield  {title} {\enquote {\bibinfo {title} {Dual negative
  differential of heat generation in a strongly correlated quantum dot
  side-coupled to {M}ajorana bound states},}\ }\href@noop {} {\bibfield
  {journal} {\bibinfo  {journal} {Front. Phys.}\ }\textbf {\bibinfo {volume}
  {9}},\ \bibinfo {pages} {727934} (\bibinfo {year} {2021})}\BibitemShut
  {NoStop}%
\bibitem [{\citenamefont {Smirnov}(2018)}]{Smirnov_2018}%
  \BibitemOpen
  \bibfield  {author} {\bibinfo {author} {\bibfnamefont {S.}~\bibnamefont
  {Smirnov}},\ }\bibfield  {title} {\enquote {\bibinfo {title} {Universal
  {M}ajorana thermoelectric noise},}\ }\href@noop {} {\bibfield  {journal}
  {\bibinfo  {journal} {Phys.\ Rev.\ B}\ }\textbf {\bibinfo {volume} {97}},\
  \bibinfo {pages} {165434} (\bibinfo {year} {2018})}\BibitemShut {NoStop}%
\bibitem [{\citenamefont {Smirnov}(2019{\natexlab{b}})}]{Smirnov_2019a}%
  \BibitemOpen
  \bibfield  {author} {\bibinfo {author} {\bibfnamefont {S.}~\bibnamefont
  {Smirnov}},\ }\bibfield  {title} {\enquote {\bibinfo {title} {Dynamic
  {M}ajorana resonances and universal symmetry of nonequilibrium thermoelectric
  quantum noise},}\ }\href@noop {} {\bibfield  {journal} {\bibinfo  {journal}
  {Phys.\ Rev.\ B}\ }\textbf {\bibinfo {volume} {100}},\ \bibinfo {pages}
  {245410} (\bibinfo {year} {2019}{\natexlab{b}})}\BibitemShut {NoStop}%
\bibitem [{\citenamefont {Smirnov}(2020)}]{Smirnov_2020a}%
  \BibitemOpen
  \bibfield  {author} {\bibinfo {author} {\bibfnamefont {S.}~\bibnamefont
  {Smirnov}},\ }\bibfield  {title} {\enquote {\bibinfo {title} {Dual {M}ajorana
  universality in thermally induced nonequilibrium},}\ }\href@noop {}
  {\bibfield  {journal} {\bibinfo  {journal} {Phys.\ Rev.\ B}\ }\textbf
  {\bibinfo {volume} {101}},\ \bibinfo {pages} {125417} (\bibinfo {year}
  {2020})}\BibitemShut {NoStop}%
\bibitem [{\citenamefont {Wrze{\'s}niewski}\ and\ \citenamefont
  {Weymann}(2021)}]{Wrzesniewski_2021}%
  \BibitemOpen
  \bibfield  {author} {\bibinfo {author} {\bibfnamefont {K.}~\bibnamefont
  {Wrze{\'s}niewski}}\ and\ \bibinfo {author} {\bibfnamefont {I.}~\bibnamefont
  {Weymann}},\ }\bibfield  {title} {\enquote {\bibinfo {title} {Magnetization
  dynamics in a {M}ajorana-wire-quantum-dot setup},}\ }\href@noop {} {\bibfield
   {journal} {\bibinfo  {journal} {Phys.\ Rev.\ B}\ }\textbf {\bibinfo {volume}
  {103}},\ \bibinfo {pages} {125413} (\bibinfo {year} {2021})}\BibitemShut
  {NoStop}%
\bibitem [{\citenamefont {Smirnov}(2015)}]{Smirnov_2015}%
  \BibitemOpen
  \bibfield  {author} {\bibinfo {author} {\bibfnamefont {S.}~\bibnamefont
  {Smirnov}},\ }\bibfield  {title} {\enquote {\bibinfo {title} {Majorana
  tunneling entropy},}\ }\href@noop {} {\bibfield  {journal} {\bibinfo
  {journal} {Phys.\ Rev.\ B}\ }\textbf {\bibinfo {volume} {92}},\ \bibinfo
  {pages} {195312} (\bibinfo {year} {2015})}\BibitemShut {NoStop}%
\bibitem [{\citenamefont {Hartman}\ \emph {et~al.}(2018)\citenamefont
  {Hartman}, \citenamefont {Olsen}, \citenamefont {L{\"u}scher}, \citenamefont
  {Samani}, \citenamefont {Fallahi}, \citenamefont {Gardner}, \citenamefont
  {Manfra},\ and\ \citenamefont {Folk}}]{Hartman_2018}%
  \BibitemOpen
  \bibfield  {author} {\bibinfo {author} {\bibfnamefont {N.}~\bibnamefont
  {Hartman}}, \bibinfo {author} {\bibfnamefont {C.}~\bibnamefont {Olsen}},
  \bibinfo {author} {\bibfnamefont {S.}~\bibnamefont {L{\"u}scher}}, \bibinfo
  {author} {\bibfnamefont {M.}~\bibnamefont {Samani}}, \bibinfo {author}
  {\bibfnamefont {S.}~\bibnamefont {Fallahi}}, \bibinfo {author} {\bibfnamefont
  {G.~C.}\ \bibnamefont {Gardner}}, \bibinfo {author} {\bibfnamefont
  {M.}~\bibnamefont {Manfra}}, \ and\ \bibinfo {author} {\bibfnamefont
  {J.}~\bibnamefont {Folk}},\ }\bibfield  {title} {\enquote {\bibinfo {title}
  {Direct entropy measurement in a mesoscopic quantum system},}\ }\href@noop {}
  {\bibfield  {journal} {\bibinfo  {journal} {Nat. Phys.}\ }\textbf {\bibinfo
  {volume} {14}},\ \bibinfo {pages} {1083} (\bibinfo {year}
  {2018})}\BibitemShut {NoStop}%
\bibitem [{\citenamefont {Kleeorin}\ \emph {et~al.}(2019)\citenamefont
  {Kleeorin}, \citenamefont {Thierschmann}, \citenamefont {Buhmann},
  \citenamefont {Georges}, \citenamefont {Molenkamp},\ and\ \citenamefont
  {Meir}}]{Kleeorin_2019}%
  \BibitemOpen
  \bibfield  {author} {\bibinfo {author} {\bibfnamefont {Y.}~\bibnamefont
  {Kleeorin}}, \bibinfo {author} {\bibfnamefont {H.}~\bibnamefont
  {Thierschmann}}, \bibinfo {author} {\bibfnamefont {H.}~\bibnamefont
  {Buhmann}}, \bibinfo {author} {\bibfnamefont {A.}~\bibnamefont {Georges}},
  \bibinfo {author} {\bibfnamefont {L.~W.}\ \bibnamefont {Molenkamp}}, \ and\
  \bibinfo {author} {\bibfnamefont {Y.}~\bibnamefont {Meir}},\ }\bibfield
  {title} {\enquote {\bibinfo {title} {How to measure the entropy of a
  mesoscopic system via thermoelectric transport},}\ }\href@noop {} {\bibfield
  {journal} {\bibinfo  {journal} {Nat. Commun.}\ }\textbf {\bibinfo {volume}
  {10}},\ \bibinfo {pages} {5801} (\bibinfo {year} {2019})}\BibitemShut
  {NoStop}%
\bibitem [{\citenamefont {Pyurbeeva}\ and\ \citenamefont
  {Mol}(2021)}]{Pyurbeeva_2021}%
  \BibitemOpen
  \bibfield  {author} {\bibinfo {author} {\bibfnamefont {E.}~\bibnamefont
  {Pyurbeeva}}\ and\ \bibinfo {author} {\bibfnamefont {J.~A.}\ \bibnamefont
  {Mol}},\ }\bibfield  {title} {\enquote {\bibinfo {title} {A thermodynamic
  approach to measuring entropy in a few-electron nanodevice},}\ }\href@noop {}
  {\bibfield  {journal} {\bibinfo  {journal} {Entropy}\ }\textbf {\bibinfo
  {volume} {23}},\ \bibinfo {pages} {640} (\bibinfo {year} {2021})}\BibitemShut
  {NoStop}%
\bibitem [{\citenamefont {Child}\ \emph
  {et~al.}(2021{\natexlab{a}})\citenamefont {Child}, \citenamefont {Sheekey},
  \citenamefont {L{\"u}scher}, \citenamefont {Fallahi}, \citenamefont
  {Gardner}, \citenamefont {Manfra}, \citenamefont {Kleeorin}, \citenamefont
  {Meir},\ and\ \citenamefont {Folk}}]{Child_2021}%
  \BibitemOpen
  \bibfield  {author} {\bibinfo {author} {\bibfnamefont {T.}~\bibnamefont
  {Child}}, \bibinfo {author} {\bibfnamefont {O.}~\bibnamefont {Sheekey}},
  \bibinfo {author} {\bibfnamefont {S.}~\bibnamefont {L{\"u}scher}}, \bibinfo
  {author} {\bibfnamefont {S.}~\bibnamefont {Fallahi}}, \bibinfo {author}
  {\bibfnamefont {G.~C.}\ \bibnamefont {Gardner}}, \bibinfo {author}
  {\bibfnamefont {M.}~\bibnamefont {Manfra}}, \bibinfo {author} {\bibfnamefont
  {Y.}~\bibnamefont {Kleeorin}}, \bibinfo {author} {\bibfnamefont
  {Y.}~\bibnamefont {Meir}}, \ and\ \bibinfo {author} {\bibfnamefont
  {J.}~\bibnamefont {Folk}},\ }\bibfield  {title} {\enquote {\bibinfo {title}
  {Entropy measurement of a strongly correlated quantum dot},}\ }\href@noop {}
  {\bibfield  {journal} {\bibinfo  {journal} {arXiv:2110.14158}\ } (\bibinfo
  {year} {2021}{\natexlab{a}})}\BibitemShut {NoStop}%
\bibitem [{\citenamefont {Child}\ \emph
  {et~al.}(2021{\natexlab{b}})\citenamefont {Child}, \citenamefont {Sheekey},
  \citenamefont {L{\"u}scher}, \citenamefont {Fallahi}, \citenamefont
  {Gardner}, \citenamefont {Manfra},\ and\ \citenamefont {Folk}}]{Child_2021a}%
  \BibitemOpen
  \bibfield  {author} {\bibinfo {author} {\bibfnamefont {T.}~\bibnamefont
  {Child}}, \bibinfo {author} {\bibfnamefont {O.}~\bibnamefont {Sheekey}},
  \bibinfo {author} {\bibfnamefont {S.}~\bibnamefont {L{\"u}scher}}, \bibinfo
  {author} {\bibfnamefont {S.}~\bibnamefont {Fallahi}}, \bibinfo {author}
  {\bibfnamefont {G.~C.}\ \bibnamefont {Gardner}}, \bibinfo {author}
  {\bibfnamefont {M.}~\bibnamefont {Manfra}}, \ and\ \bibinfo {author}
  {\bibfnamefont {J.}~\bibnamefont {Folk}},\ }\bibfield  {title} {\enquote
  {\bibinfo {title} {A robust protocol for entropy measurement in mesoscopic
  circuits},}\ }\href@noop {} {\bibfield  {journal} {\bibinfo  {journal}
  {arXiv:2110.14172}\ } (\bibinfo {year} {2021}{\natexlab{b}})}\BibitemShut
  {NoStop}%
\bibitem [{\citenamefont {Sela}\ \emph {et~al.}(2019)\citenamefont {Sela},
  \citenamefont {Oreg}, \citenamefont {Plugge}, \citenamefont {Hartman},
  \citenamefont {L{\"u}scher},\ and\ \citenamefont {Folk}}]{Sela_2019}%
  \BibitemOpen
  \bibfield  {author} {\bibinfo {author} {\bibfnamefont {E.}~\bibnamefont
  {Sela}}, \bibinfo {author} {\bibfnamefont {Y.}~\bibnamefont {Oreg}}, \bibinfo
  {author} {\bibfnamefont {S.}~\bibnamefont {Plugge}}, \bibinfo {author}
  {\bibfnamefont {N.}~\bibnamefont {Hartman}}, \bibinfo {author} {\bibfnamefont
  {S.}~\bibnamefont {L{\"u}scher}}, \ and\ \bibinfo {author} {\bibfnamefont
  {J.}~\bibnamefont {Folk}},\ }\bibfield  {title} {\enquote {\bibinfo {title}
  {Detecting the universal fractional entropy of {M}ajorana zero modes},}\
  }\href@noop {} {\bibfield  {journal} {\bibinfo  {journal} {Phys.\ Rev.\
  Lett.}\ }\textbf {\bibinfo {volume} {123}},\ \bibinfo {pages} {147702}
  (\bibinfo {year} {2019})}\BibitemShut {NoStop}%
\bibitem [{\citenamefont {Smirnov}(2021)}]{Smirnov_2021}%
  \BibitemOpen
  \bibfield  {author} {\bibinfo {author} {\bibfnamefont {S.}~\bibnamefont
  {Smirnov}},\ }\bibfield  {title} {\enquote {\bibinfo {title} {{M}ajorana
  entropy revival via tunneling phases},}\ }\href@noop {} {\bibfield  {journal}
  {\bibinfo  {journal} {Phys.\ Rev.\ B}\ }\textbf {\bibinfo {volume} {103}},\
  \bibinfo {pages} {075440} (\bibinfo {year} {2021})}\BibitemShut {NoStop}%
\bibitem [{\citenamefont {Gong}\ \emph {et~al.}(2021)\citenamefont {Gong},
  \citenamefont {Dai}, \citenamefont {Zhang}, \citenamefont {Jiang},\ and\
  \citenamefont {Gong}}]{Gong_2021}%
  \BibitemOpen
  \bibfield  {author} {\bibinfo {author} {\bibfnamefont {T.}~\bibnamefont
  {Gong}}, \bibinfo {author} {\bibfnamefont {X.-F.}\ \bibnamefont {Dai}},
  \bibinfo {author} {\bibfnamefont {L.-L.}\ \bibnamefont {Zhang}}, \bibinfo
  {author} {\bibfnamefont {C.}~\bibnamefont {Jiang}}, \ and\ \bibinfo {author}
  {\bibfnamefont {W.-J.}\ \bibnamefont {Gong}},\ }\bibfield  {title} {\enquote
  {\bibinfo {title} {Interference effect on the {A}ndreev reflections induced
  by {M}ajorana bound states},}\ }\href@noop {} {\bibfield  {journal} {\bibinfo
   {journal} {J. Phys. Condens. Matter}\ }\textbf {\bibinfo {volume} {33}},\
  \bibinfo {pages} {215303} (\bibinfo {year} {2021})}\BibitemShut {NoStop}%
\bibitem [{\citenamefont {Gau}\ \emph {et~al.}(2020{\natexlab{a}})\citenamefont
  {Gau}, \citenamefont {Egger}, \citenamefont {Zazunov},\ and\ \citenamefont
  {Gefen}}]{Gau_2020}%
  \BibitemOpen
  \bibfield  {author} {\bibinfo {author} {\bibfnamefont {M.}~\bibnamefont
  {Gau}}, \bibinfo {author} {\bibfnamefont {R.}~\bibnamefont {Egger}}, \bibinfo
  {author} {\bibfnamefont {A.}~\bibnamefont {Zazunov}}, \ and\ \bibinfo
  {author} {\bibfnamefont {Y.}~\bibnamefont {Gefen}},\ }\bibfield  {title}
  {\enquote {\bibinfo {title} {Towards dark space stabilization and
  manipulation in driven dissipative {M}ajorana platforms},}\ }\href@noop {}
  {\bibfield  {journal} {\bibinfo  {journal} {Phys.\ Rev.\ B}\ }\textbf
  {\bibinfo {volume} {102}},\ \bibinfo {pages} {134501} (\bibinfo {year}
  {2020}{\natexlab{a}})}\BibitemShut {NoStop}%
\bibitem [{\citenamefont {Gau}\ \emph {et~al.}(2020{\natexlab{b}})\citenamefont
  {Gau}, \citenamefont {Egger}, \citenamefont {Zazunov},\ and\ \citenamefont
  {Gefen}}]{Gau_2020a}%
  \BibitemOpen
  \bibfield  {author} {\bibinfo {author} {\bibfnamefont {M.}~\bibnamefont
  {Gau}}, \bibinfo {author} {\bibfnamefont {R.}~\bibnamefont {Egger}}, \bibinfo
  {author} {\bibfnamefont {A.}~\bibnamefont {Zazunov}}, \ and\ \bibinfo
  {author} {\bibfnamefont {Yuval}\ \bibnamefont {Gefen}},\ }\bibfield  {title}
  {\enquote {\bibinfo {title} {Driven dissipative {M}ajorana dark spaces},}\
  }\href@noop {} {\bibfield  {journal} {\bibinfo  {journal} {Phys.\ Rev.\
  Lett.}\ }\textbf {\bibinfo {volume} {125}},\ \bibinfo {pages} {147701}
  (\bibinfo {year} {2020}{\natexlab{b}})}\BibitemShut {NoStop}%
\bibitem [{\citenamefont {Smirnov}\ and\ \citenamefont
  {Grifoni}(2011{\natexlab{a}})}]{Smirnov_2011a}%
  \BibitemOpen
  \bibfield  {author} {\bibinfo {author} {\bibfnamefont {S.}~\bibnamefont
  {Smirnov}}\ and\ \bibinfo {author} {\bibfnamefont {M.}~\bibnamefont
  {Grifoni}},\ }\bibfield  {title} {\enquote {\bibinfo {title} {Slave-boson
  {K}eldysh field theory for the {K}ondo effect in quantum dots},}\ }\href@noop
  {} {\bibfield  {journal} {\bibinfo  {journal} {Phys.\ Rev.\ B}\ }\textbf
  {\bibinfo {volume} {84}},\ \bibinfo {pages} {125303} (\bibinfo {year}
  {2011}{\natexlab{a}})}\BibitemShut {NoStop}%
\bibitem [{\citenamefont {Smirnov}\ and\ \citenamefont
  {Grifoni}(2011{\natexlab{b}})}]{Smirnov_2011b}%
  \BibitemOpen
  \bibfield  {author} {\bibinfo {author} {\bibfnamefont {S.}~\bibnamefont
  {Smirnov}}\ and\ \bibinfo {author} {\bibfnamefont {M.}~\bibnamefont
  {Grifoni}},\ }\bibfield  {title} {\enquote {\bibinfo {title} {Kondo effect in
  interacting nanoscopic systems: {K}eldysh field integral theory},}\
  }\href@noop {} {\bibfield  {journal} {\bibinfo  {journal} {Phys.\ Rev.\ B}\
  }\textbf {\bibinfo {volume} {84}},\ \bibinfo {pages} {235314} (\bibinfo
  {year} {2011}{\natexlab{b}})}\BibitemShut {NoStop}%
\bibitem [{\citenamefont {Niklas}\ \emph {et~al.}(2016)\citenamefont {Niklas},
  \citenamefont {Smirnov}, \citenamefont {Mantelli}, \citenamefont
  {Marga{\'n}ska}, \citenamefont {Nguyen}, \citenamefont {Wernsdorfer},
  \citenamefont {Cleuziou},\ and\ \citenamefont {Grifoni}}]{Niklas_2016}%
  \BibitemOpen
  \bibfield  {author} {\bibinfo {author} {\bibfnamefont {M.}~\bibnamefont
  {Niklas}}, \bibinfo {author} {\bibfnamefont {S.}~\bibnamefont {Smirnov}},
  \bibinfo {author} {\bibfnamefont {D.}~\bibnamefont {Mantelli}}, \bibinfo
  {author} {\bibfnamefont {M.}~\bibnamefont {Marga{\'n}ska}}, \bibinfo {author}
  {\bibfnamefont {N.-V.}\ \bibnamefont {Nguyen}}, \bibinfo {author}
  {\bibfnamefont {W.}~\bibnamefont {Wernsdorfer}}, \bibinfo {author}
  {\bibfnamefont {J.-P.}\ \bibnamefont {Cleuziou}}, \ and\ \bibinfo {author}
  {\bibfnamefont {M.}~\bibnamefont {Grifoni}},\ }\bibfield  {title} {\enquote
  {\bibinfo {title} {Blocking transport resonances via {K}ondo many-body
  entanglement in quantum dots},}\ }\href@noop {} {\bibfield  {journal}
  {\bibinfo  {journal} {Nat. Commun.}\ }\textbf {\bibinfo {volume} {7}},\
  \bibinfo {pages} {12442} (\bibinfo {year} {2016})}\BibitemShut {NoStop}%
\bibitem [{\citenamefont {Altland}\ and\ \citenamefont
  {Simons}(2010)}]{Altland_2010}%
  \BibitemOpen
  \bibfield  {author} {\bibinfo {author} {\bibfnamefont {A.}~\bibnamefont
  {Altland}}\ and\ \bibinfo {author} {\bibfnamefont {B.}~\bibnamefont
  {Simons}},\ }\href@noop {} {\emph {\bibinfo {title} {Condensed Matter Field
  Theory}}},\ \bibinfo {edition} {2nd}\ ed.\ (\bibinfo  {publisher} {Cambridge
  University Press, Cambridge},\ \bibinfo {year} {2010})\BibitemShut {NoStop}%
\bibitem [{\citenamefont {Meir}\ and\ \citenamefont
  {Wingreen}(1992)}]{Meir_1992}%
  \BibitemOpen
  \bibfield  {author} {\bibinfo {author} {\bibfnamefont {Y.}~\bibnamefont
  {Meir}}\ and\ \bibinfo {author} {\bibfnamefont {N.~S.}\ \bibnamefont
  {Wingreen}},\ }\bibfield  {title} {\enquote {\bibinfo {title} {Landauer
  formula for the current through an interacting electron region},}\
  }\href@noop {} {\bibfield  {journal} {\bibinfo  {journal} {Phys.\ Rev.\
  Lett.}\ }\textbf {\bibinfo {volume} {68}},\ \bibinfo {pages} {2512} (\bibinfo
  {year} {1992})}\BibitemShut {NoStop}%
\bibitem [{\citenamefont {Li}\ and\ \citenamefont {Xu}(2020)}]{Li_2020}%
  \BibitemOpen
  \bibfield  {author} {\bibinfo {author} {\bibfnamefont {X.-Q.}\ \bibnamefont
  {Li}}\ and\ \bibinfo {author} {\bibfnamefont {L.}~\bibnamefont {Xu}},\
  }\bibfield  {title} {\enquote {\bibinfo {title} {Nonlocality of {M}ajorana
  zero modes and teleportation: {S}elf-consistent treatment based on the
  {B}ogoliubov-de {G}ennes equation},}\ }\href@noop {} {\bibfield  {journal}
  {\bibinfo  {journal} {Phys.\ Rev.\ B}\ }\textbf {\bibinfo {volume} {101}},\
  \bibinfo {pages} {205401} (\bibinfo {year} {2020})}\BibitemShut {NoStop}%
\bibitem [{\citenamefont {Ren}\ \emph {et~al.}(2021)\citenamefont {Ren},
  \citenamefont {Ke}, \citenamefont {Guo}, \citenamefont {Zhang},\ and\
  \citenamefont {L{\"u}}}]{Ren_2021}%
  \BibitemOpen
  \bibfield  {author} {\bibinfo {author} {\bibfnamefont {J.-T.}\ \bibnamefont
  {Ren}}, \bibinfo {author} {\bibfnamefont {S.-S.}\ \bibnamefont {Ke}},
  \bibinfo {author} {\bibfnamefont {Y.}~\bibnamefont {Guo}}, \bibinfo {author}
  {\bibfnamefont {H.-W.}\ \bibnamefont {Zhang}}, \ and\ \bibinfo {author}
  {\bibfnamefont {H.-F.}\ \bibnamefont {L{\"u}}},\ }\bibfield  {title}
  {\enquote {\bibinfo {title} {Phase diagram and quantum transport in a
  semiconductor-superconductor hybrid nanowire with long-range pairing
  interactions},}\ }\href@noop {} {\bibfield  {journal} {\bibinfo  {journal}
  {Phys.\ Rev.\ B}\ }\textbf {\bibinfo {volume} {103}},\ \bibinfo {pages}
  {045428} (\bibinfo {year} {2021})}\BibitemShut {NoStop}%
\bibitem [{\citenamefont {Leumer}\ \emph {et~al.}(2021)\citenamefont {Leumer},
  \citenamefont {Grifoni}, \citenamefont {Muralidharan},\ and\ \citenamefont
  {Marganska}}]{Leumer_2021}%
  \BibitemOpen
  \bibfield  {author} {\bibinfo {author} {\bibfnamefont {N.}~\bibnamefont
  {Leumer}}, \bibinfo {author} {\bibfnamefont {M.}~\bibnamefont {Grifoni}},
  \bibinfo {author} {\bibfnamefont {B.}~\bibnamefont {Muralidharan}}, \ and\
  \bibinfo {author} {\bibfnamefont {M.}~\bibnamefont {Marganska}},\ }\bibfield
  {title} {\enquote {\bibinfo {title} {Linear and nonlinear transport across a
  finite kitaev chain: {A}n exact analytical study},}\ }\href@noop {}
  {\bibfield  {journal} {\bibinfo  {journal} {Phys.\ Rev.\ B}\ }\textbf
  {\bibinfo {volume} {103}},\ \bibinfo {pages} {165432} (\bibinfo {year}
  {2021})}\BibitemShut {NoStop}%
\bibitem [{\citenamefont {Mourik}\ \emph {et~al.}(2012)\citenamefont {Mourik},
  \citenamefont {Zuo}, \citenamefont {Frolov}, \citenamefont {Plissard},
  \citenamefont {Bakkers},\ and\ \citenamefont {Kouwenhoven}}]{Mourik_2012}%
  \BibitemOpen
  \bibfield  {author} {\bibinfo {author} {\bibfnamefont {V.}~\bibnamefont
  {Mourik}}, \bibinfo {author} {\bibfnamefont {K.}~\bibnamefont {Zuo}},
  \bibinfo {author} {\bibfnamefont {S.~M.}\ \bibnamefont {Frolov}}, \bibinfo
  {author} {\bibfnamefont {S.~R.}\ \bibnamefont {Plissard}}, \bibinfo {author}
  {\bibfnamefont {E.~P. A.~M.}\ \bibnamefont {Bakkers}}, \ and\ \bibinfo
  {author} {\bibfnamefont {L.~P.}\ \bibnamefont {Kouwenhoven}},\ }\bibfield
  {title} {\enquote {\bibinfo {title} {Signatures of {M}ajorana fermions in
  hybrid superconductor-semiconductor nanowire devices},}\ }\href@noop {}
  {\bibfield  {journal} {\bibinfo  {journal} {Science}\ }\textbf {\bibinfo
  {volume} {336}},\ \bibinfo {pages} {1003} (\bibinfo {year}
  {2012})}\BibitemShut {NoStop}%
\end{thebibliography}
\end{document}